\newif\ifPREPRINT
\PREPRINTtrue

\ifPREPRINT
  \documentclass[a4paper]{article}
  \usepackage{authblk}
\else
  \documentclass[authoryear]{elsarticle}
\fi
\usepackage{amsfonts}
\usepackage{amsmath}
\usepackage{graphicx}
\usepackage{hyperref}
\usepackage[utf8]{inputenc}
\usepackage[T1]{fontenc}
\usepackage{natbib}
\usepackage{units}

\newcommand{\Abstract}{%
  \begin{abstract}
    Wear on total knee replacements (TKRs) is an important criterion for their performance characteristics.
    Numerical simulations of such wear has seen increasing attention over the last years.
    They have the potential to be much faster and less expensive than the in vitro tests in use today.  While it is
    unlikely that in silico tests will replace actual physical tests in the foreseeable future, a judicious combination
    of both approaches can help making both implant design and pre-clinical testing quicker and more cost-effective.

    The challenge today for the design of simulation methods is to obtain results that convey quantitative information,
    and to do so quickly and reliably.
    This involves the choice of mathematical models as well as the numerical tools used to solve
    them.  The correctness of the choice can only be validated by comparing with experimental results.

    In this paper we present finite element simulations of the wear in TKRs during the gait cycle standardized in
    the ISO~14243-1 document, used for compliance testing in several countries.  As the ISO~14243-1 standard
    is precisely defined and publicly available, it can serve as an excellent benchmark for comparison
    of wear simulation methods.
    Our novel contact algorithm works without Lagrange multipliers and penalty methods, achieving
    unparalleled stability and efficiency.
    We compare our simulation results with the experimental data from physical tests using two different actual
    TKRs, each test being performed three times.
    We can closely predict the total mass loss due to wear after five million gait cycles.
    We also observe a good match between the wear patterns seen in experiments and our simulation results.
  \end{abstract}
}

\begin{document}

\ifPREPRINT
  \title{Simulating Wear \\ On Total Knee Replacements}
  \author[1]{Ansgar Burchardt\footnote{email: \href{mailto:ansgar.burchardt@tu-dresden.de}{ansgar.burchardt@tu-dresden.de}}}
  \author[2]{Christian Abicht\footnote{email: \href{mailto:christian.abicht@questmed.de}{christian.abicht@questmed.de}}}
  \author[1]{Oliver Sander\footnote{email: \href{mailto:oliver.sander@tu-dresden.de}{oliver.sander@tu-dresden.de}}}
  \affil[1]{Technische Universität Dresden, Dresden, Germany}
  \affil[2]{Questmed GmbH, Kleinmachnow, Germany}

  \maketitle
  \Abstract
\fi


\section{Introduction}

The wear between tibial plateau and femur component is one of the main limiting factors for the life-span of total knee replacements (TKRs).
In the course of millions of gait cycles, the hard femur head grinds off small particles from the tibial plateau, which is usually made from relatively soft polyethylene (UHMWPE).
Small microparticles start to migrate within the knee joint, leading to inflammation and eventually osteolysis \citep{Gupta2007}.
In extreme cases, mechanical failure of (the surface of) the tibial plateau is observed.

To limit these risks various national guidelines require pre-clinical in vitro testing of the wear behavior of knee implants.
These tests are performed by knee wear testing machines.
The precise conditions are formulated in a series of documents published by the International Standards
Organisation.  We focus here on ISO~14243-1~\citep{ISO14243-1}, which describes testing conditions
of a load-controlled testing gait cycle for normal walking.
A wear test consists of five million such cycles, and of monitoring the mass loss
of the tibial bearing component.

Performing such experimental tests is a cost-intensive task.
The required five million cycles take about three months of time.
Sometimes tests have to be aborted in mid way, because the initial positioning was not well chosen.
(Indeed, the initial positioning of the femur and tibia with respect to each other is not precisely specified by ISO~14243-1.)
Its determination remains an important open problem even for the designers of knee implants.

Computer simulations of these standardized tests can help to reduce costs and time-to-market.
While simulations cannot replace the actual compliance tests, they can help to avoid some of the preliminary tests that need to be performed during the design phase of a new implant.
In particular, numerical simulations can help to determine suitable initial configurations.
With this information, the number of actual physical experiments is greatly reduced.

For these reasons, the numerical simulation of TKR wear behavior has seen increasing interest over the
past years.  Various finite element and rigid body models appear in the literature, all combining different contact
formulations and wear laws.  While some authors consider Archard's wear law to be sufficient
\citep{OBrien2013,Willing2009}, others focus on developing more advanced laws to better capture
the behavior of UHMWPE \citep{Abdelgaied2011, OBrien2014}.  Several groups use the ISO~14243 test family
as a benchmark problem \citep{OBrien2012, Willing2009}, but only the latter group uses the load-controlled
variant 14243-1. \cite{Abdelgaied2011} and \cite{OBrien2014} compare their findings with experimental results.

In this contribution we describe a new finite element model of two TKRs including the
tibia plateau and femoral component. Following \cite{Abicht2005,Willing2009},
we model both components as deformable objects, because numerical tests showed that surprisingly
little run-time can be saved by keeping the femur rigid.
We model the contact between the two objects exactly, with a surface--to--surface (mortar)
discretization~\citep{Wohlmuth2011,laursen:2002} without
recurse to any regularization parameter.  The wear on the tibial plateau is described using Archard's
wear law.  We compare the predicted wear patterns and total wear mass loss to values obtained
by experimental testing, and observe a very good correspondence.  In particular, the choice of the
comparatively simple Archard's law appears to be appropriate.  More complicated models may be useful
in the future, but need more experimental data to be justified.

Numerical wear testing involving deformable objects requires the solution of many contact problems,
in particular if long-time wear, statistical effects, or shape-optimization is involved.
Most articles mentioned above use commercial finite element software, which typically
uses Lagrange multipliers or penalty approaches for the contact problems.  The well-known
drawbacks are additional degrees of freedom, artificial parameters, penalization errors,
and instabilities.  In contrast, our contact model uses a novel nonsmooth multigrid algorithm
which solves the contact problems directly.  This avoids all abovementioned
drawbacks---instead, the solver is provably convergent, and as fast as solvers for linear
elastic problems without contact \citep{graeser_sack_sander:2009}.
We implemented the model in a C++ code based on the open-source \textsc{Dune} library (\url{www.dune-project.org}).
The efficiency and stability of our algorithm allows to solve challenging wear simulations
quickly and reliably.

\section{Methods}

We simulate the ISO~14243-1 testing cycle using a finite element model that includes both the femur component and the tibial inlay of the implant as deformable bodies.
The two interact by a contact condition which we model using a surface--to--surface (mortar) contact
discretization~\citep{laursen:2002,Wohlmuth2011}.
As TKRs are well lubricated inside the testing machine we assume the contact to be frictionless.
The numerical code is our own research implementation based on the open-source finite element code
\textsc{Dune}~\citep{DUNE24,bastian_buse_sander:2010} (\url{www.dune-project.org}).

\subsection{Finite Element Model}

\begin{table}
  \centering
  \begin{tabular}{rr|rr}
    mesh && \# elements & \# vertices \\
    \hline
    Mebio 1
         & femur & 13\,824 & 4\,633 \\
         & tibia & 3\,948 & 1\,205 \\
    Mebio 2
         & femur & 31\,432 & 9\,234 \\
         & tibia & 81\,400 & 17\,445 \\
    Mebio 3
         & femur & 184\,526 & 42\,750 \\
         & tibia & 188\,993 & 37\,982 \\
    \hline
    Genius Pro 1
         & femur & 9\,238 & 2\,820 \\
         & tibia & 4\,720 & 1\,496 \\
    Genius Pro 2
         & femur & 33\,864 & 8\,745 \\
         & tibia & 18\,799 & 5\,038 \\
    Genius Pro 3
         & femur & 109\,045 & 24\,550 \\
         & tibia & 121\,548 & 26\,484
  \end{tabular}
  \caption{Mesh resolutions.  For each of the two TKRs, meshes in three different resolutions
           were constructed.}
  \label{tab:meshes}
\end{table}

We tested our simulation procedure with two implants which are sold commercially, and for which all geometry data and the original experimental results of the ISO~14243-1 compliance testing were gratefully made available to us
by the manufacturer (aap Implantate GmbH).

These are a ``Mebio 2015'' (in the following: Mebio \citep{QuestMed2015})
and a ``Genius Pro (Fixed Bearing, PCL retaining)'' (in the following Genius Pro \citep{Endolab2011}).
CAD data of the TKR volumes was available in Parasolid format.
Tetrahedral volume meshes with different resolutions were constructed using ANSYS (ANSYS Inc., Canonsburg, PA)
and the open source mesh generator \textsc{gmsh}~\citep{Geuzaine2009}.
The meshes for the tibial component have between 5\,000 and 120\,000 elements, and 1\,500 to 26\,500 vertices
for the Genius Pro implant, and between 4\,000 and 190\,000 elements (1\,200 to 38\,000 vertices)
for the Mebio implant, see Table~\ref{tab:meshes}.
Four-node tetrahedral finite elements were used for the discretization.

\begin{figure}
  \centering
  \includegraphics[width=.32\textwidth]{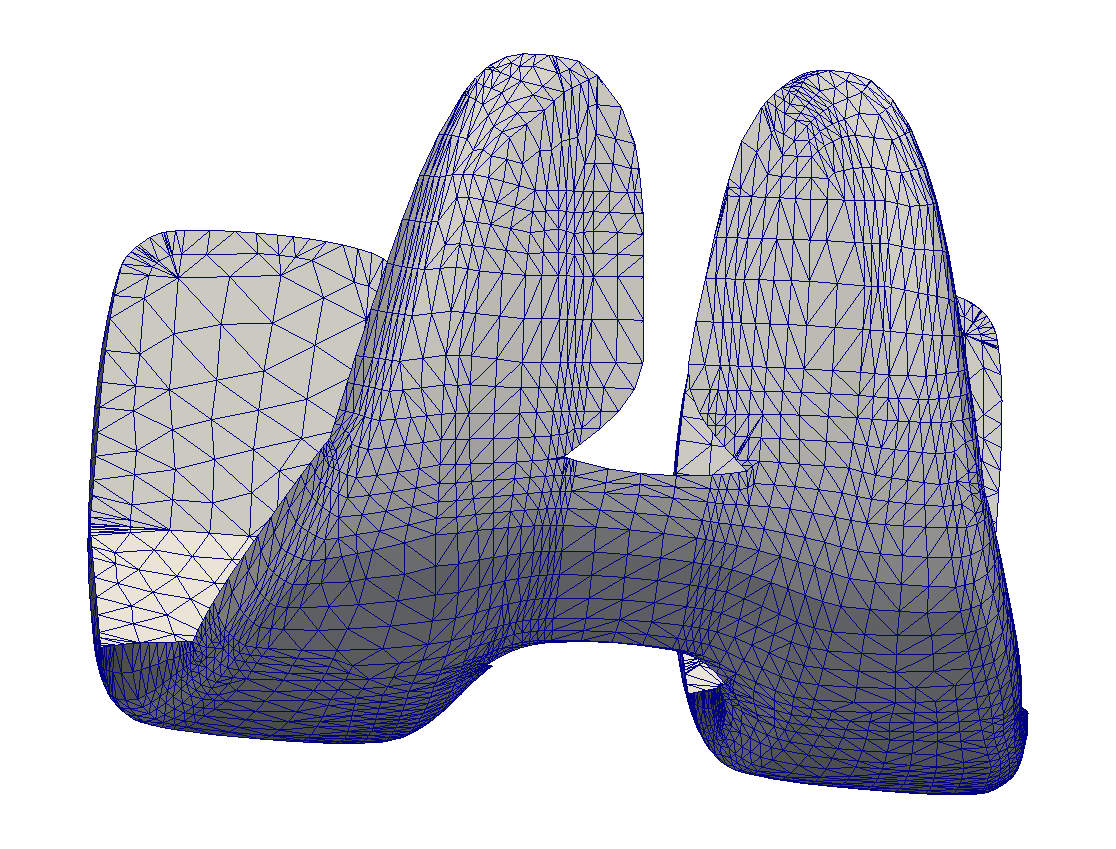} 
  \includegraphics[width=.32\textwidth]{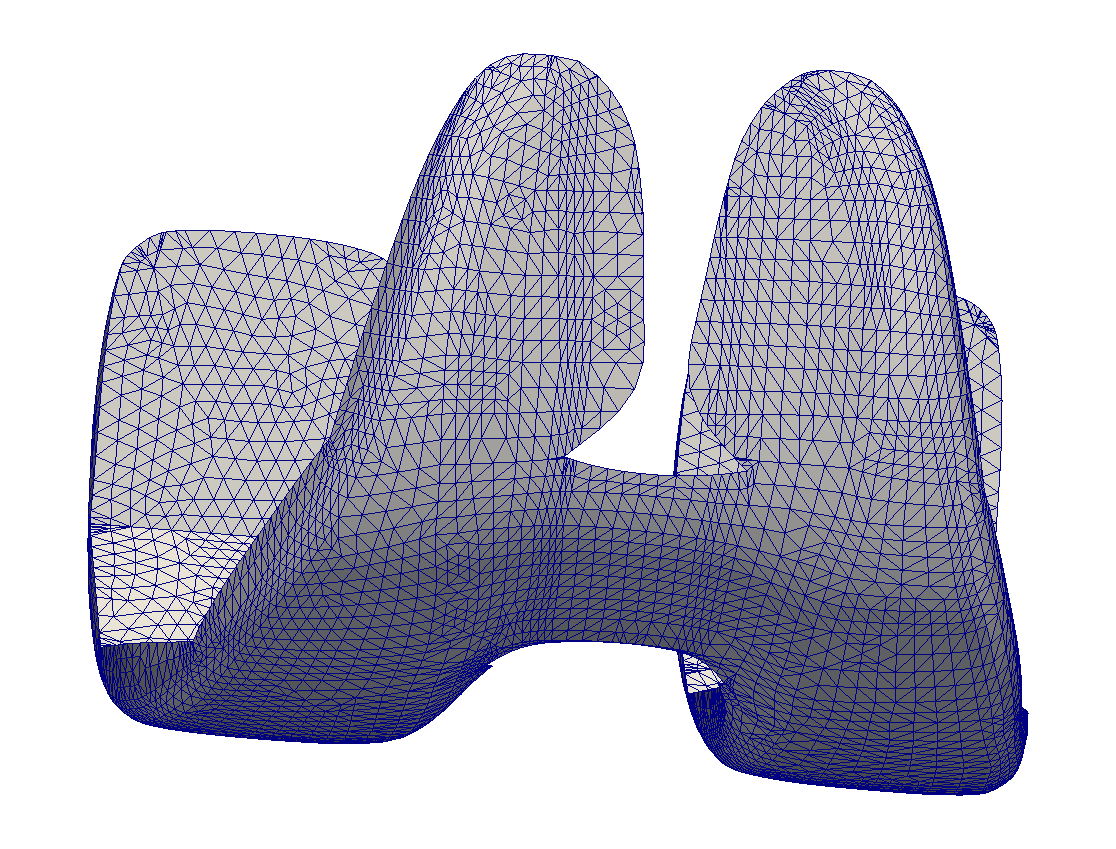} 
  \includegraphics[width=.32\textwidth]{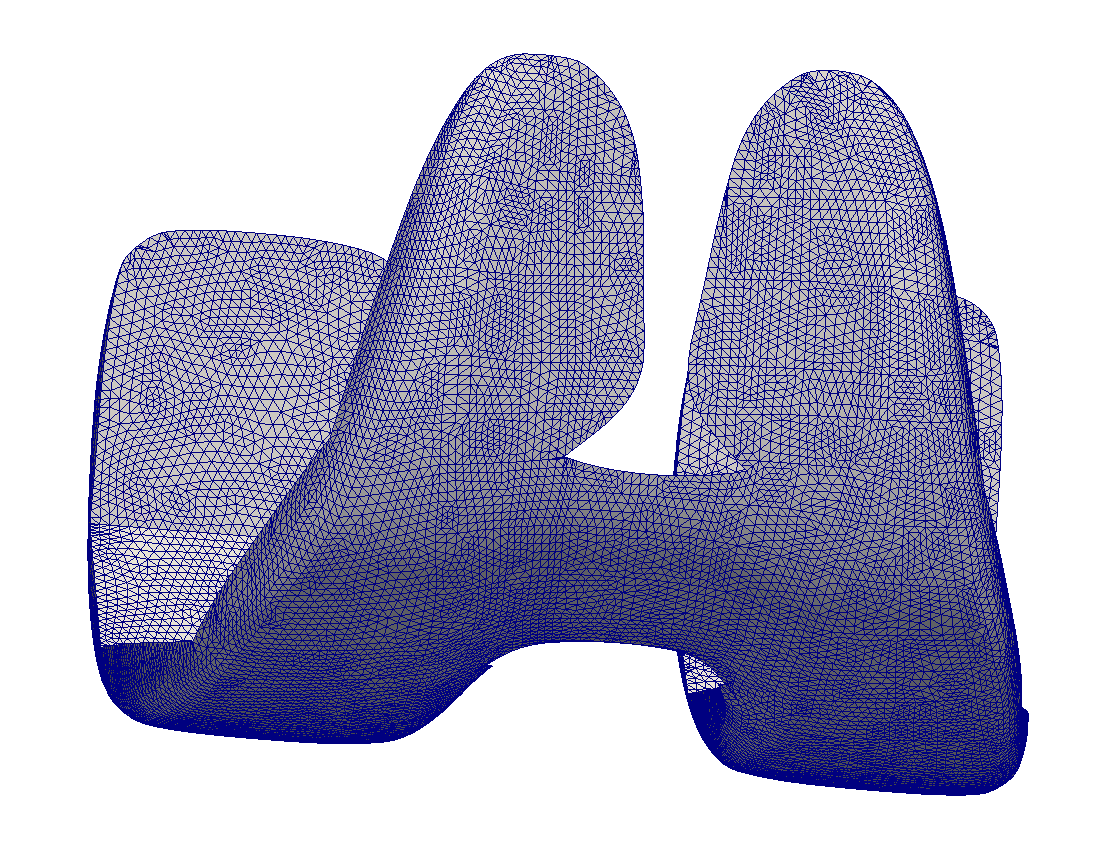} 
  \caption{Meshes for femur component of Mebio implant}
  \label{fig:meshes:mebio:femur}
\end{figure}

\begin{figure}
  \centering
  \includegraphics[width=.32\textwidth]{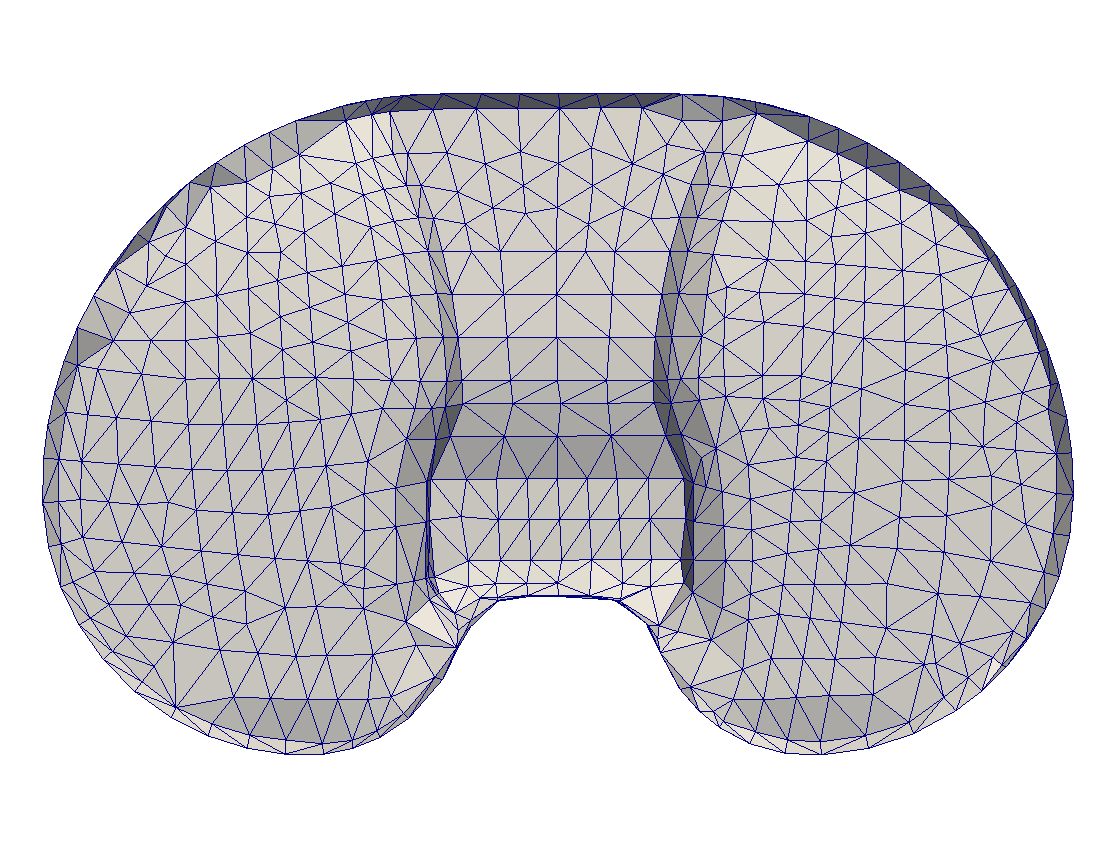} 
  \includegraphics[width=.32\textwidth]{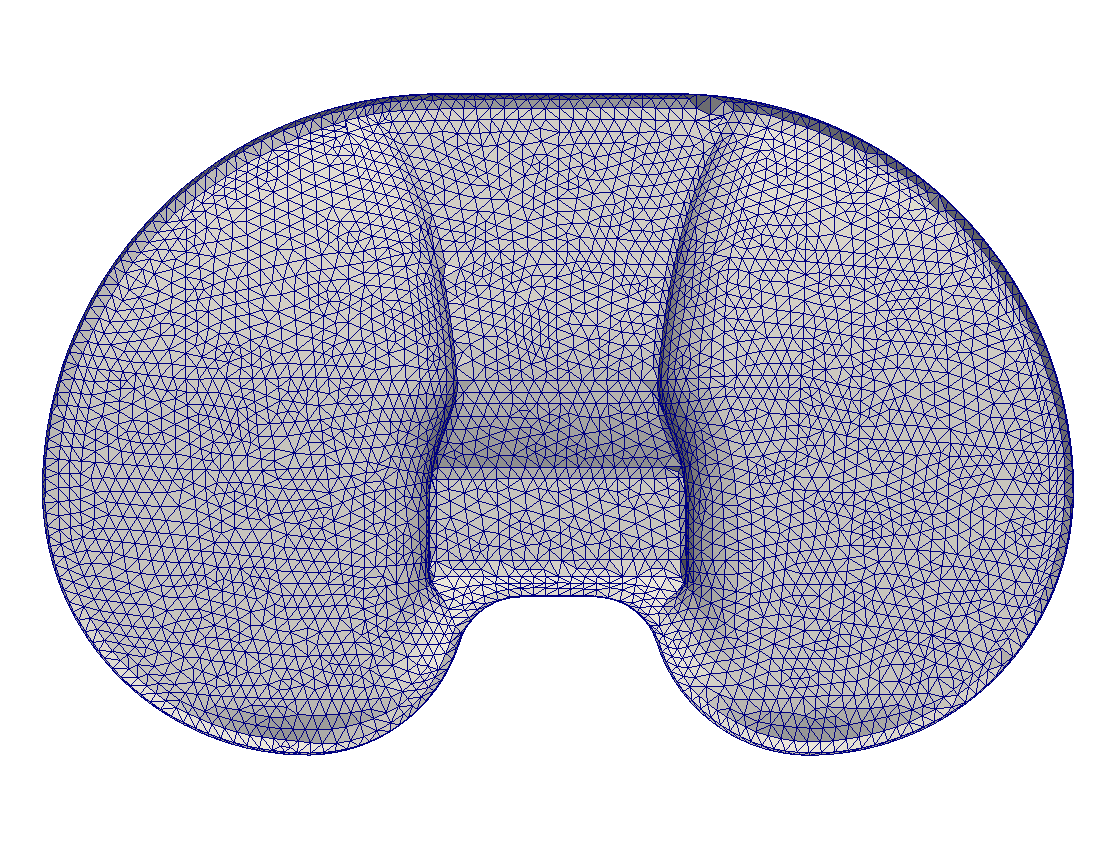} 
  \includegraphics[width=.32\textwidth]{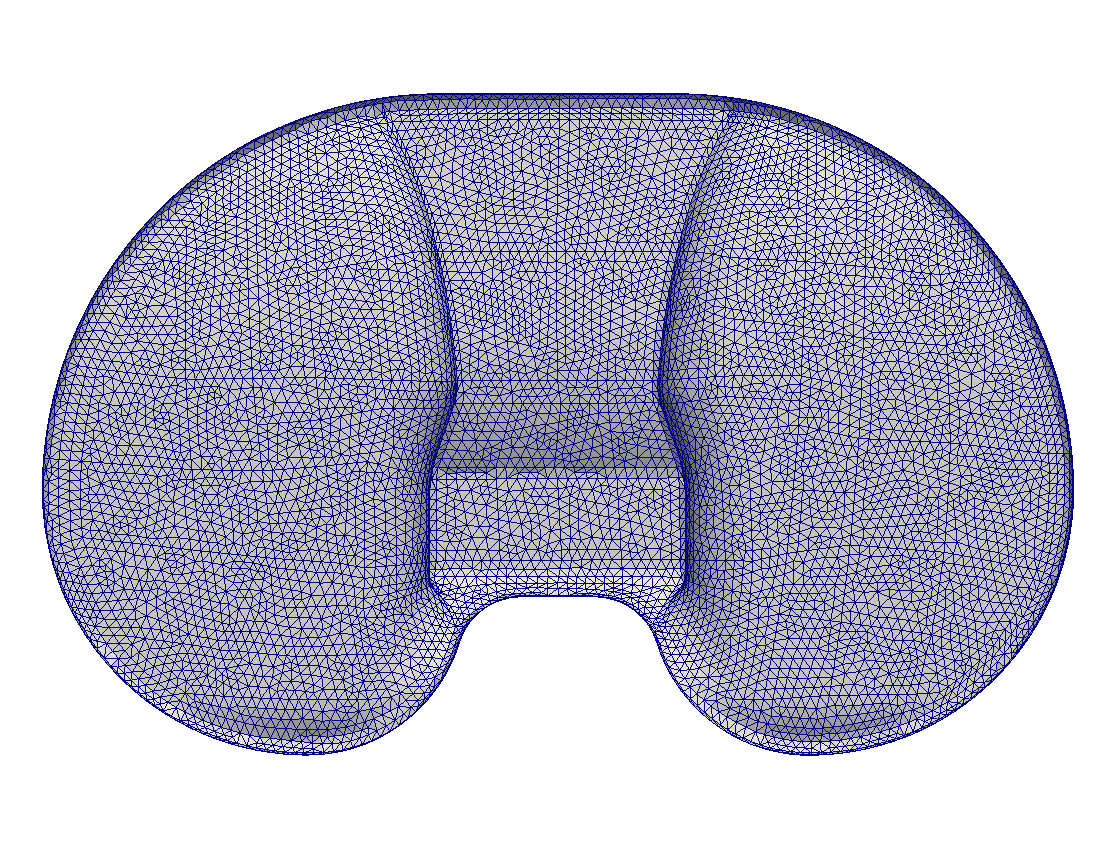} 
  \caption{Meshes for tibial component of Mebio implant}
  \label{fig:meshes:mebio:tibial}
\end{figure}

\begin{figure}
  \centering
  \includegraphics[width=.32\textwidth]{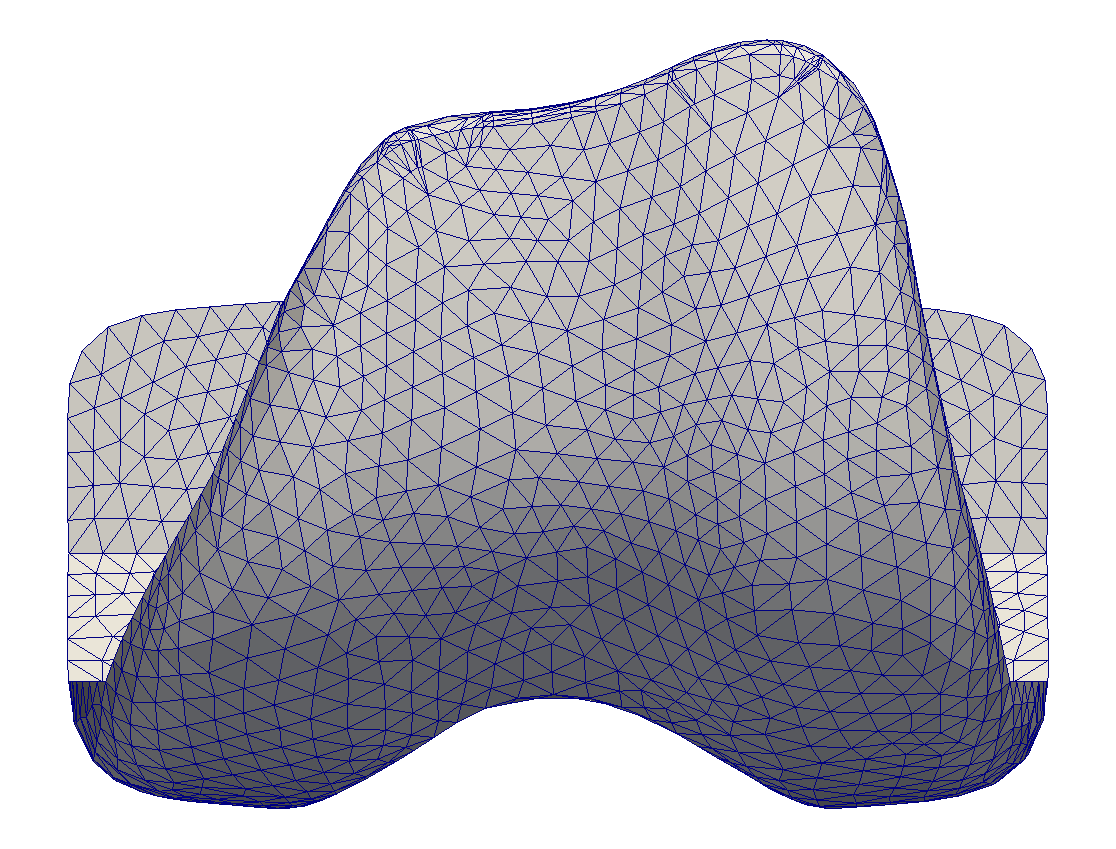} 
  \includegraphics[width=.32\textwidth]{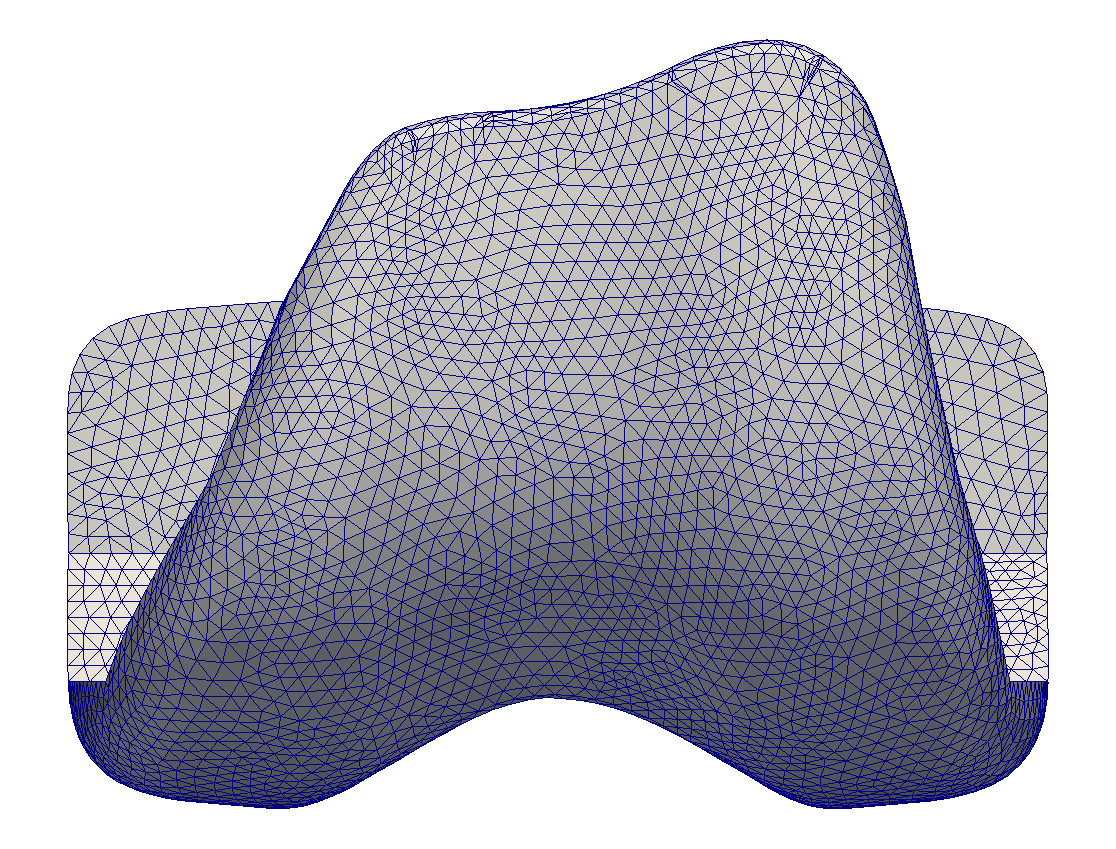} 
  \includegraphics[width=.32\textwidth]{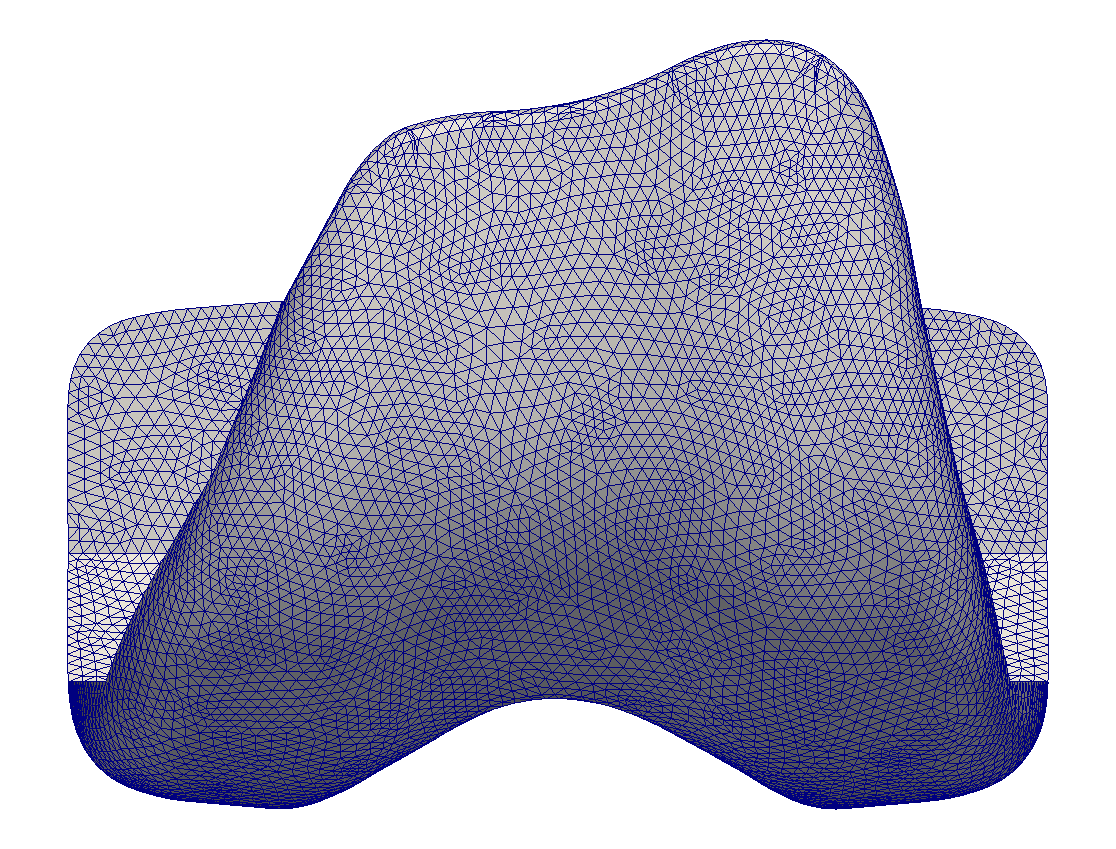} 
  \caption{Meshes for femur component of Genius Pro implant}
  \label{fig:meshes:geniuspro:femur}
\end{figure}

\begin{figure}
  \centering
  \includegraphics[width=.32\textwidth]{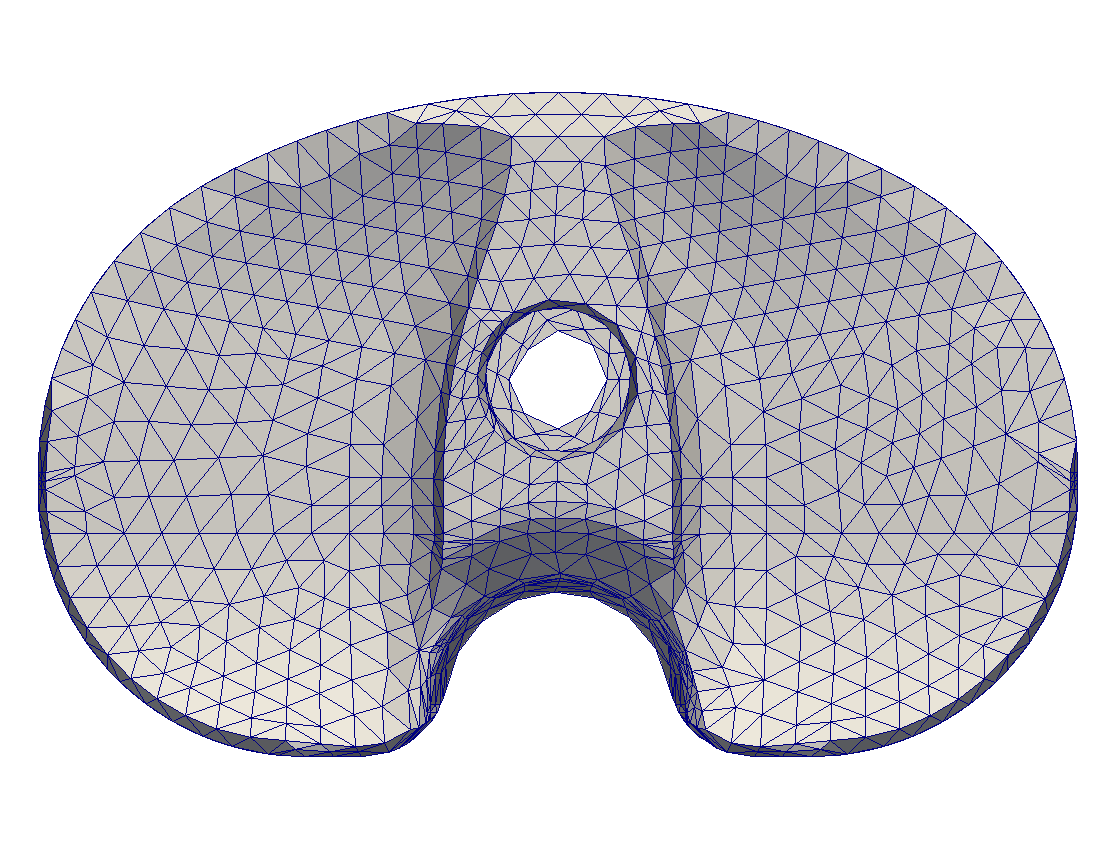} 
  \includegraphics[width=.32\textwidth]{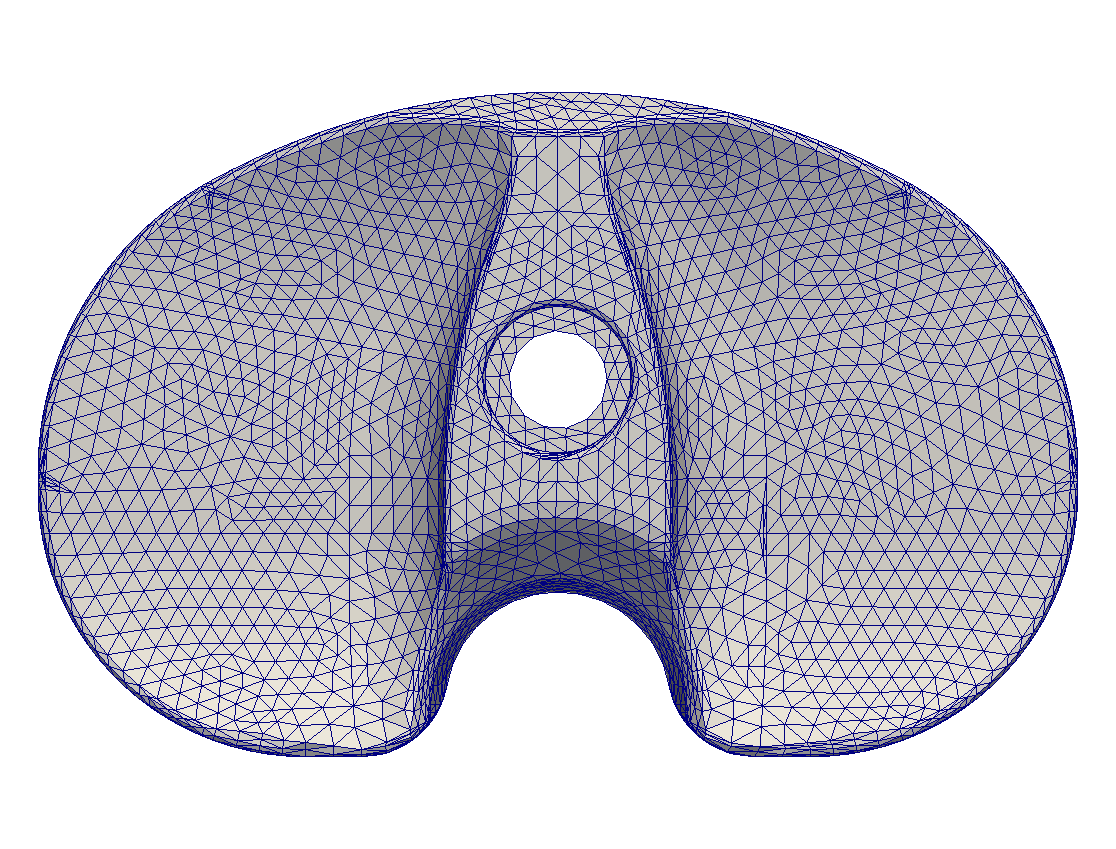} 
  \includegraphics[width=.32\textwidth]{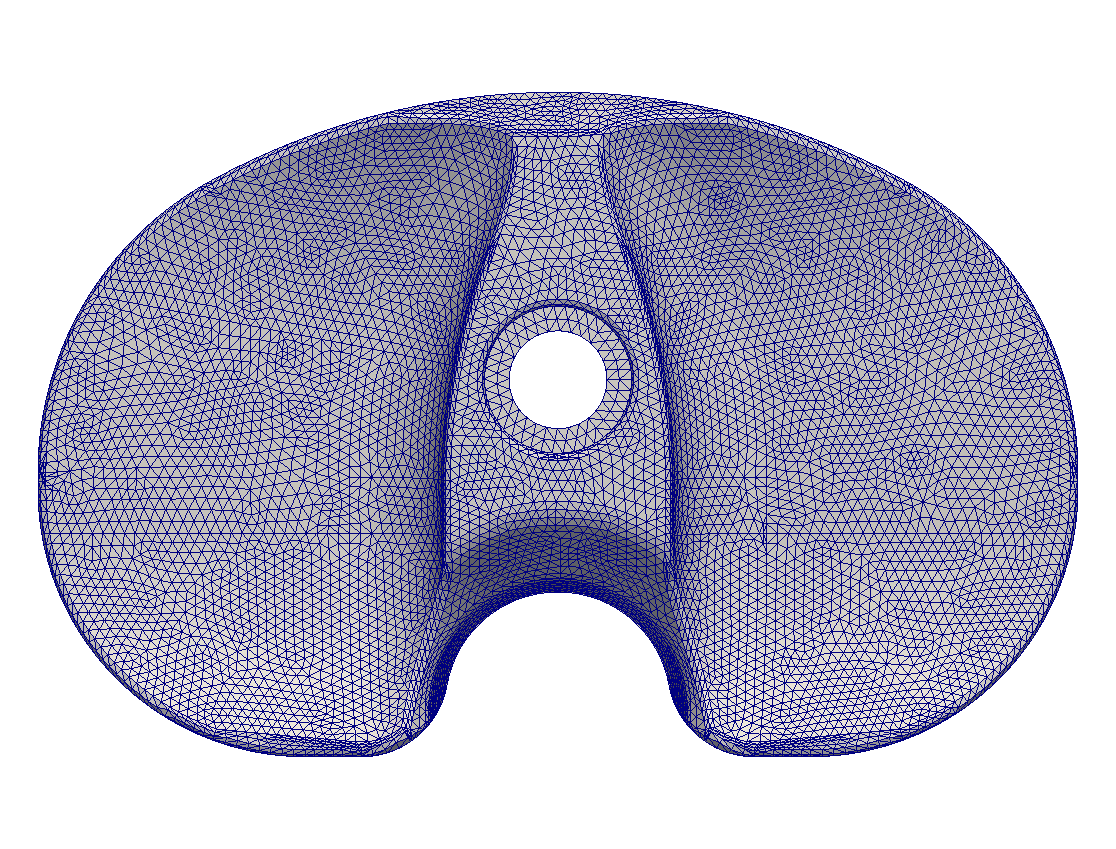} 
  \caption{Meshes for tibial component of Genius Pro implant}
  \label{fig:meshes:geniuspro:tibial}
\end{figure}

To simplify the mesh construction, the geometries of the implants was simplified slightly on their respective back sides.
The  geometric modifications did not involve areas close to the contact region.
Geometric changes this far away from the contact surface do not have a relevant influence on the wear behavior.
Figures~\ref{fig:meshes:mebio:femur} and~\ref{fig:meshes:mebio:tibial} show the meshes used for the Mebio implant;
Figures~\ref{fig:meshes:geniuspro:femur} and~\ref{fig:meshes:geniuspro:tibial} the ones used for the Genius Pro implant.

We model the UHMWPE of the tibial inlay as well as the cobalt--chromium--molybdenum alloy
of the femur component
as homogeneous, isotropic, linear elastic materials.
Material parameters are taken from~\cite[p.\,51]{Abicht2005}.
We use Young's modulus $E = 220$\,GPa, Poisson ratio $\nu = 0.31$ for the femur,
and $E = 1.1$\,GPa, $\nu = 0.42$ for the tibial inlay.

\begin{figure}
  \centering
  \includegraphics[width=.35\textwidth]{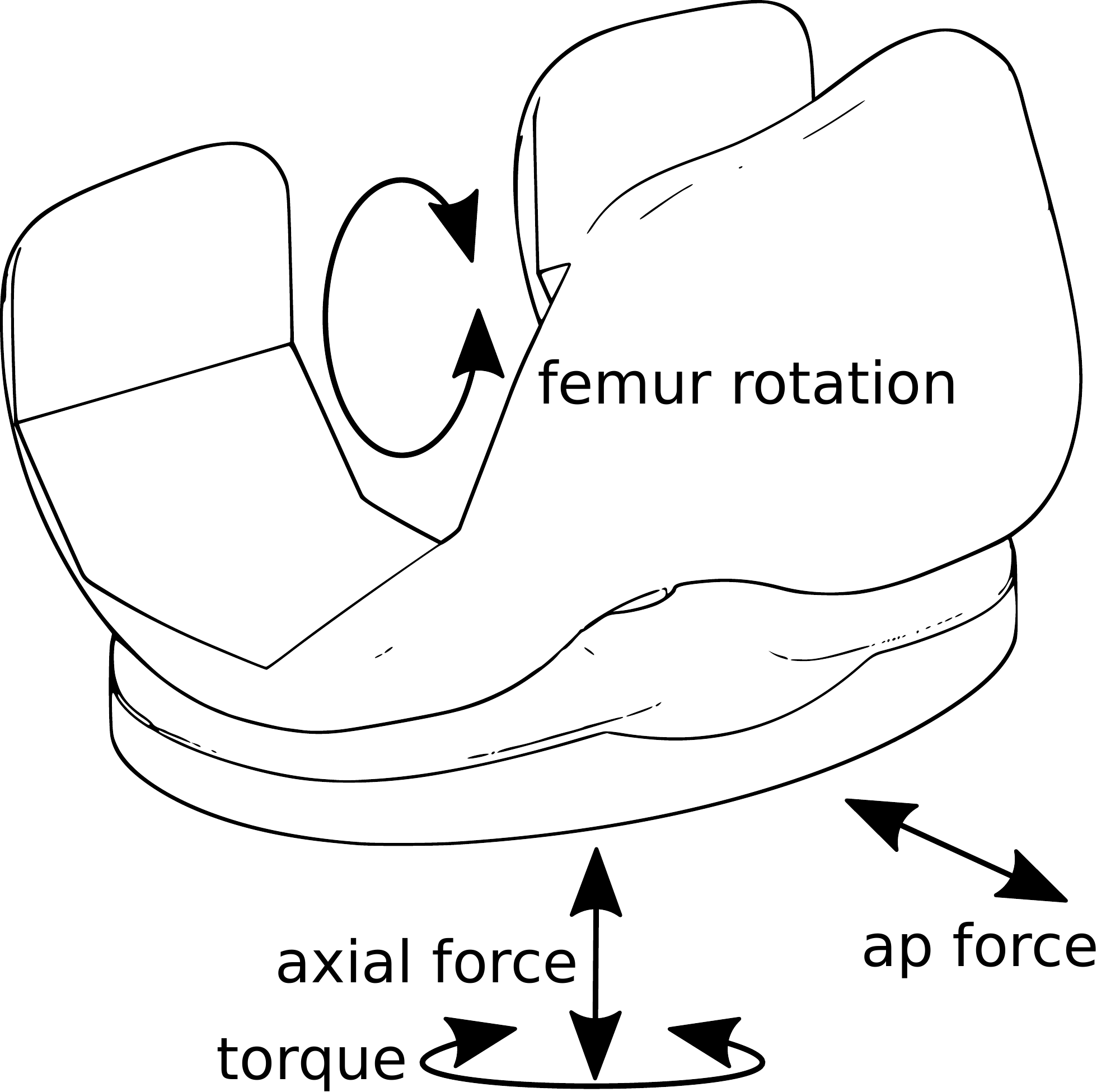}
  \hspace{0.05\textwidth}
  \includegraphics[width=.5\textwidth]{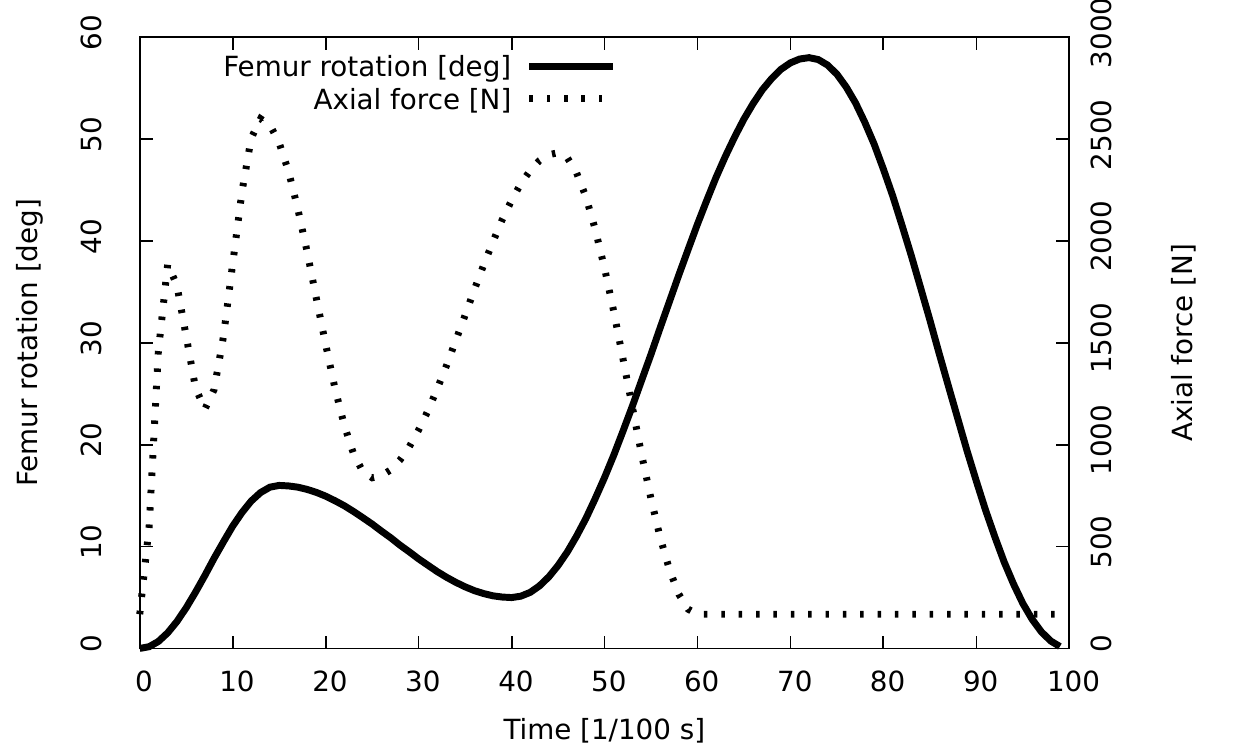}
  \caption{Boundary conditions during the gait cycle}
  \label{fig:boundaryconditions}
\end{figure}

Time-dependent boundary conditions are set as described by the \mbox{ISO~14243-1} standard.
The standard describes a single gait cycle by giving a table listing boundary values for 100 discrete
time steps within one gait cycle.
In particular, the femur component is rotated following its movement in a knee bend,
and then fixed using a displacement condition on its back side.
The boundary conditions for the tibial inlay are slightly more complicated
(Figure~\ref{fig:boundaryconditions}):
\begin{itemize}
\item The main load is an axial force pushing the tibial side upwards against the femur component,
  with time-dependent values between $\unit[167.6]{N}$ and $\unit[2600]{N}$,
\item a smaller force acts on the inferior side of the tibial inlay in anterior--posterior direction
  with values between $-\unit[265]{N}$ and $\unit[110]{N}$,
\item an internal--external torque around the axis of the axial force is applied,
  with values between $-\unit[1]{Nm}$ and $\unit[6]{Nm}$.
\end{itemize}
Additionally, the inlay is kept in place by springs penalizing anterior--posterior displacement
and internal--external rotation.
Note that this testing cycle is load-controlled and no displacement boundary condition is applied
to the tibial side at all.  This results in rank-deficient system matrices, making the finite element
system particularly challenging to solve.

For each of the 100 time steps that make up the gait cycle, we need to solve a linear contact problem.
As our model does not include inertia or rate effects, the results of these 100 steps are independent of each
other and can be computed in parallel using OpenMP.

To stay within the limits of linear elasticity we use separate reference configurations for each time step
of the gait cycle.  The displacement boundary condition describes a rigid body motion of the femur component.
We apply this motion to the femur component, and use the result as the reference configuration
for the femur.  In this rotated configuration,
the force boundary and contact conditions do not lead to large strains or displacements,
and the use of linear elasticity is justified.

We model the contact between the two objects by a penalty-free surface--to--surface (mortar) finite element discretization
as described in~\cite{Wohlmuth2003}.
Mortar discretizations avoid the unphysical stress oscillations that are known
to occur in node-to-node contact methods~\citep{Wohlmuth2011}.  The method is parameter-free,
meaning that there are no additional values that would need proper tuning to make the algorithm
function properly.  The resulting systems of equations are solved using a nonsmooth multigrid
method \citep{graeser_sack_sander:2009} together with
the interior-point solver IPOpt~\citep{Waechter2006}, which is guaranteed to converge to the solution in all cases,
without the need for artificial load stepping.

\subsection{Wear and Grid Deformation}
\label{sec:wear-and-grid-deformation}

We use Archard's wear law~\citep{Archard1956} to model the wear on the tibial inlay.
For each point on the contact surface,
Archard's law models the wear depth $l$ at that point.  The wear depth
$l$ at a time $t$ is given by
\begin{equation*}
 l(t) = k \int_0^t |p(t) v(t)| \,dt,
\end{equation*}
where $k$ is a material constant, $p$ is the contact pressure at this point,
and $v$ is the relative velocity between the two objects.
If $p$ is independent of time, and the sliding velocity is constant, then we obtain the formula
\begin{equation*}
 l = kps,
\end{equation*}
with the sliding distance $s$ typically seen in the literature.  Total volume loss is
computed as the integral of $l$ over the contact surface.
Pressure $p$ and current velocity $v$ are computed from the finite element model.
For the wear constant $k$ we chose the value $2.0\cdot10^{-7} \textrm{mm}^3/\textrm{Nm}$, taken from \citet[model M1]{OBrien2014}.
This is an important point: we did not use the experimental data available to us to determine the important
wear coefficient $k$.

\bigskip

\begin{figure}
  \centering
  \includegraphics[width=0.48\textwidth]{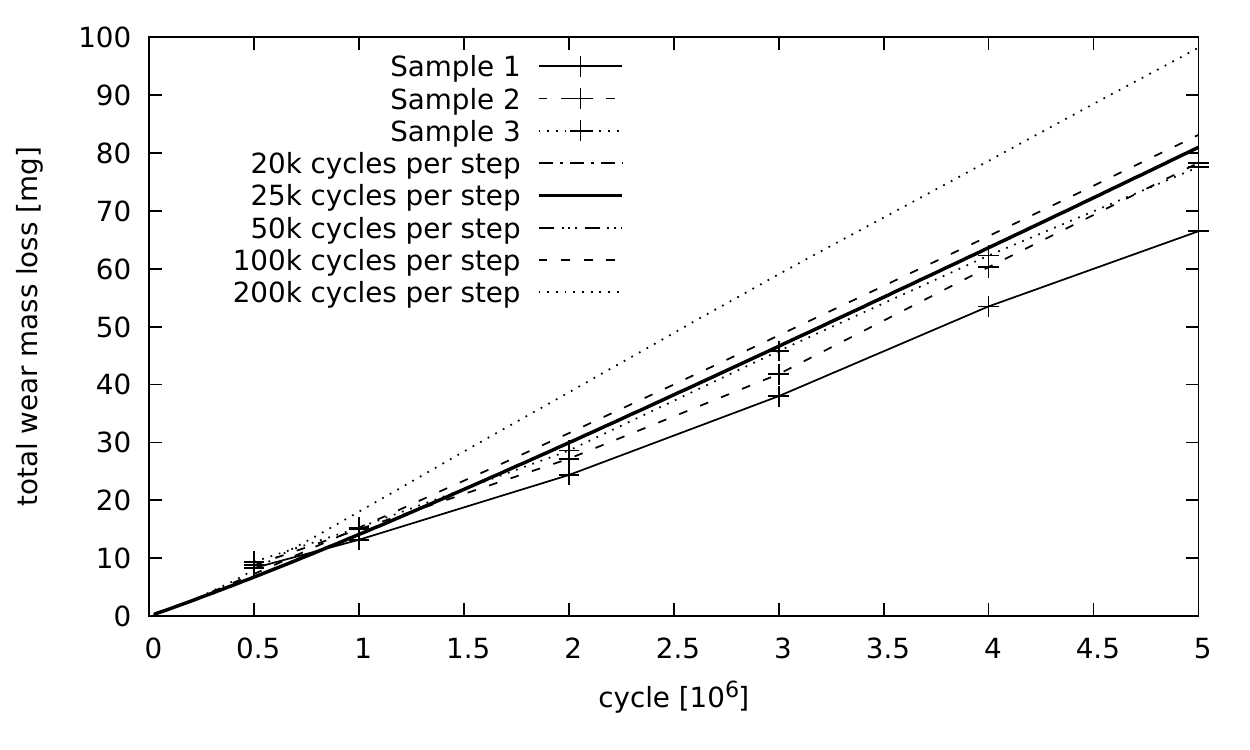}
  \includegraphics[width=0.48\textwidth]{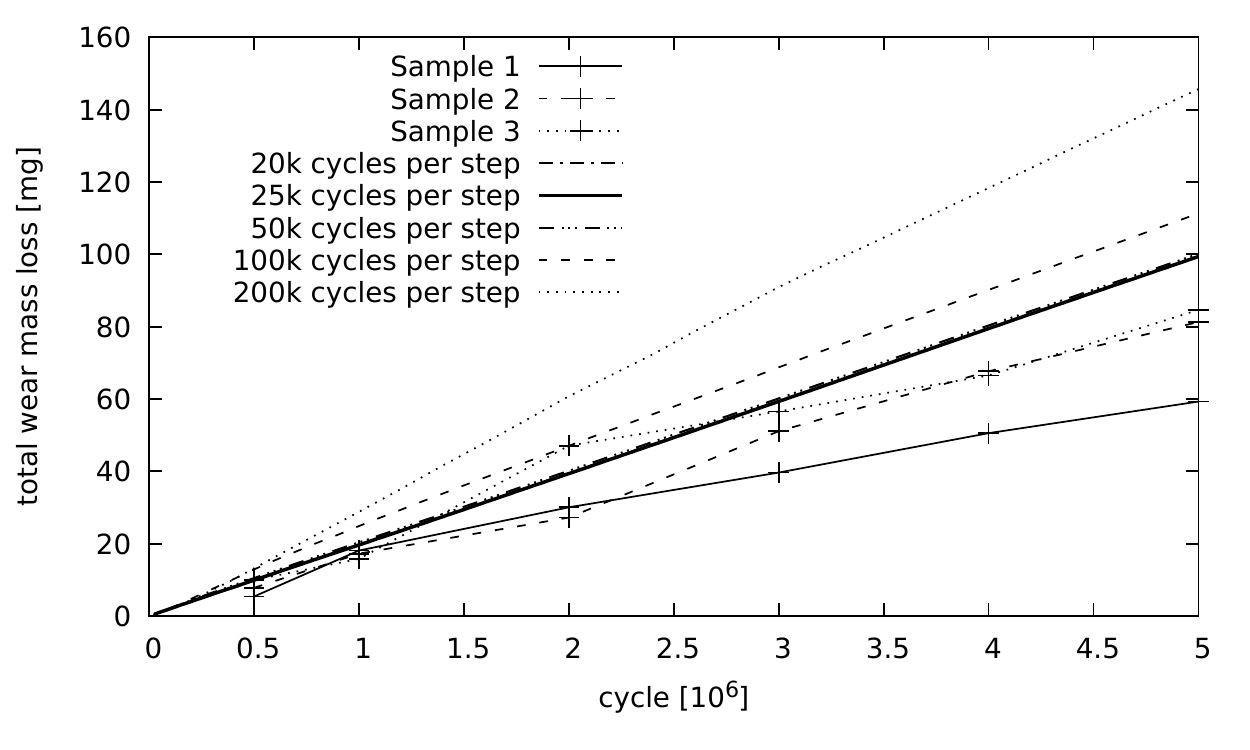}
  \caption{%
    Convergence of the total wear mass loss with smaller cycle steps for the Mebio (left)
    and Genius Pro (right) implants.
    The lines for 20k, 25k and 50k cycles per step are overlapping.
  }
  \label{fig:cycle-step-comparison}
\end{figure}

Even with a fast numerical solver, solving the 100 time steps for each of the 5~million gait cycles
mandated by ISO~14243-1 is not possible.
However, since Archard's wear law does not directly depend on the wear history, we can extrapolate the wear values
obtained for a single gait cycle over a larger number of cycles.  We
estimate the wear over $n_\Delta$ cycles by extrapolating the wear of a single gait cycle linearly,
i.e., by multiplying $l$ by $n_\Delta$.  To obtain a reasonable value for the cycle step size $n_\Delta$,
we performed simulations on a single grid per implant and compared the total wear loss over time for
different choices of $n_\Delta$ between $2\cdot10^4$ and $2\cdot10^5$.
The results from Figure~\ref{fig:cycle-step-comparison} show that a step size of $5\cdot10^4$ cycles
per step is a good choice for $n_\Delta$.

Linear extrapolation works over numbers of gait cycles that are not too large.
However, when very large numbers of cycles are considered, the change in geometry needs to be taken into account.
This change will be hardly noticable on the femur, but very relevant on the softer tibial inlay.
Therefore, after each $n_\Delta$ gait cycles, we deform the tibial grid by adding the extrapolated wear depth
in normal direction to
the grid contact surface.  From then on, this modified geometry is used as the reference configuration
for the subsequent simulation steps.
As the wear depth $l$ is small, adding it to the grid hardly influences the grid quality at all.  To be
on the safe side, after changing the grid boundary we also adjust the inner grid node positions by moving
them according to the solution of a linear elastic problem for the tibial inlay with the wear
depth $l$ as displacement boundary condition.

Combining finite element simulations, linear extrapolation, and grid deformation, we arrive at the
following hybrid time-stepping method to compute long-time wear patterns and mass loss.
Pick an extrapolation step size $n_\Delta$, and perform the following three steps:
\begin{enumerate}
\item Compute surface wear for a single gait cycle.
\item Multiply result by $n_\Delta$ to obtain the extrapolated values for $n_\Delta$ gait cycles.
\item Modify geometry using the extrapolated wear depth.
\end{enumerate}
These three steps are repeated until the total number of gait cycles reaches 5~millions.
The result is the total mass loss due to wear over time, and the wear patterns in the tibia that
can be read off directly from the grid.

\section{Results}
\label{sec:results}

\begin{figure}
  \centering
  \includegraphics[width=0.48\textwidth]{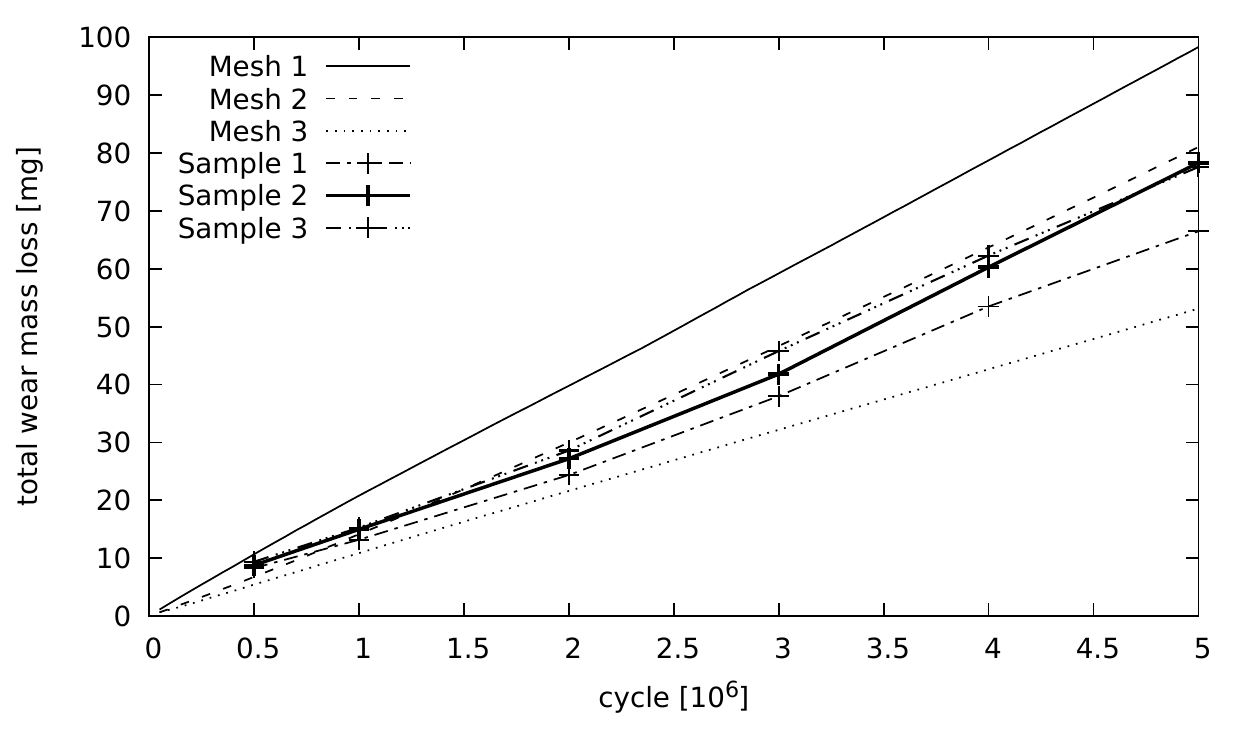}
  \includegraphics[width=0.48\textwidth]{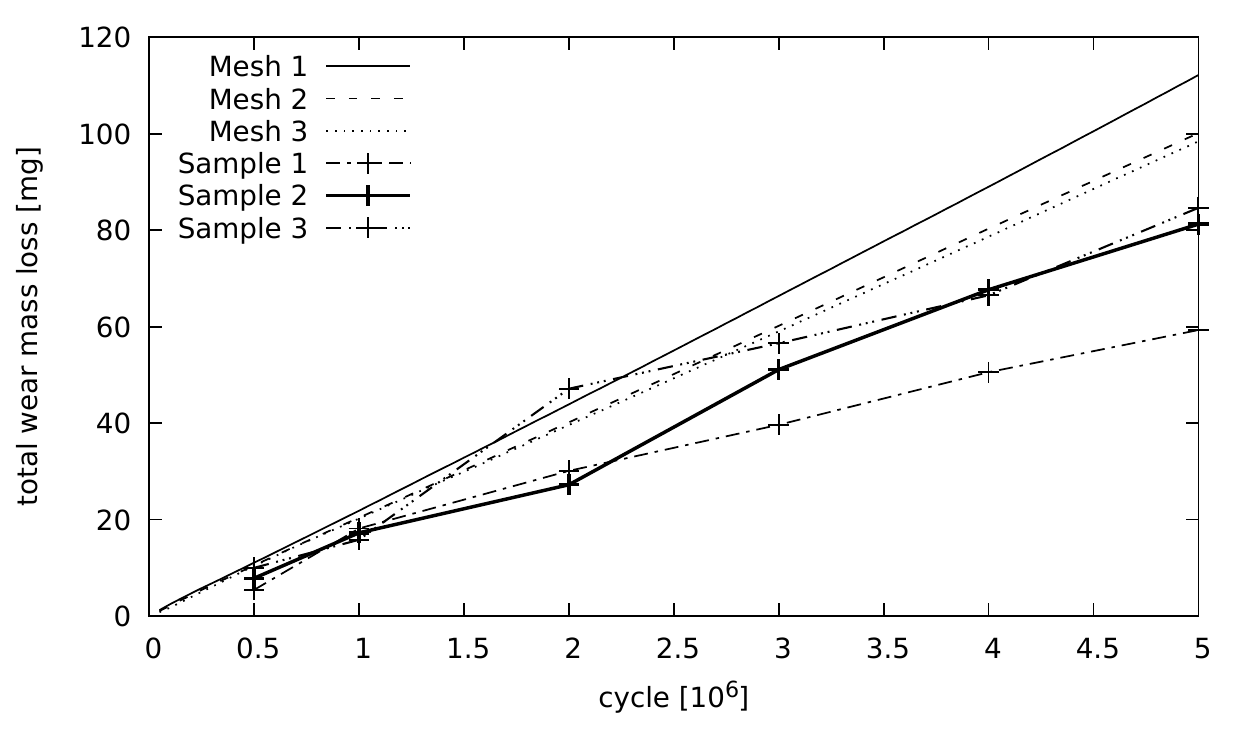}
  \caption{%
    Wear mass loss for the Mebio (left) and Genius Pro (right) implants.
    The figure shows both experimental results taken at intervals of one million gait cycles and the simulation results.
    The geometry was adapted every $5\cdot10^4$ gait cycles.
  }
  \label{fig:wear-mass-loss}
\end{figure}

The simulation results match the results of the experimental testing very well.
Figure~\ref{fig:wear-mass-loss} shows the total mass loss due to wear after up to 5~million gait cycles for the two implant geometries.
Three sets of experimental data are given for each implant, because the same in vitro experiment was repeated three
times to get an idea of the overall variance of the measurements. For both implant types there are two sets of data that
agree very well with each other, whereas the third one shows lower values.

The simulation predicts \unit[98.31/81.04/53.13]{mg} of mass loss for the Mebio for the three grids,
and \unit[112.18/100.15/98.44]{mg} for the Genius Pro after 5~million cycles.
The corresponding experimental results are in the range of \unit[66.5--77.6]{mg} for the Mebio, and between
\unit[59.33--84.65]{mg} for the Genius Pro.
We observe that mass loss for the Genius Pro implant appears to converge for increasing mesh resolution,
whereas the corresponding number for the Mebio implant does not.
This highlights the difficult nonlinear nature of the contact/wear problem.

We also observe that the experimental curves start with a steeper slope
that flattens after about 1~million gait cycles, an effect that is not reproduced by the simulation.
The presumed cause for this is that unused implants have a rougher surface, which becomes
smoother after an initial number of cycles.  Surface smoothness could be integrated easily into our model
by using an iteration-dependent wear coefficient $k$.  For simplicity, we have used the classical
Archard's law that assumes the wear constant $k$ to be fixed, because the overall effect seem to be
negligeable.

\begin{figure}
  \centering
  \includegraphics[width=0.3\textwidth]{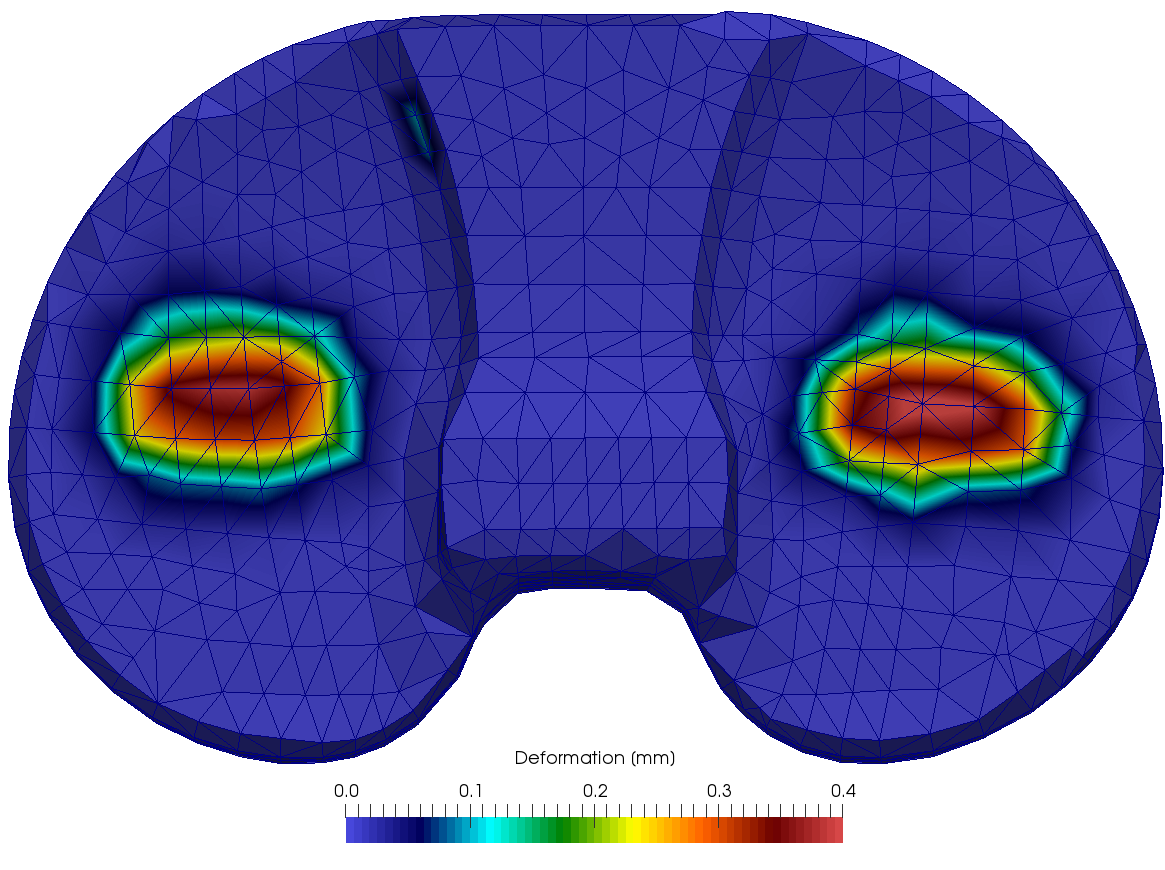}
  \includegraphics[width=0.3\textwidth]{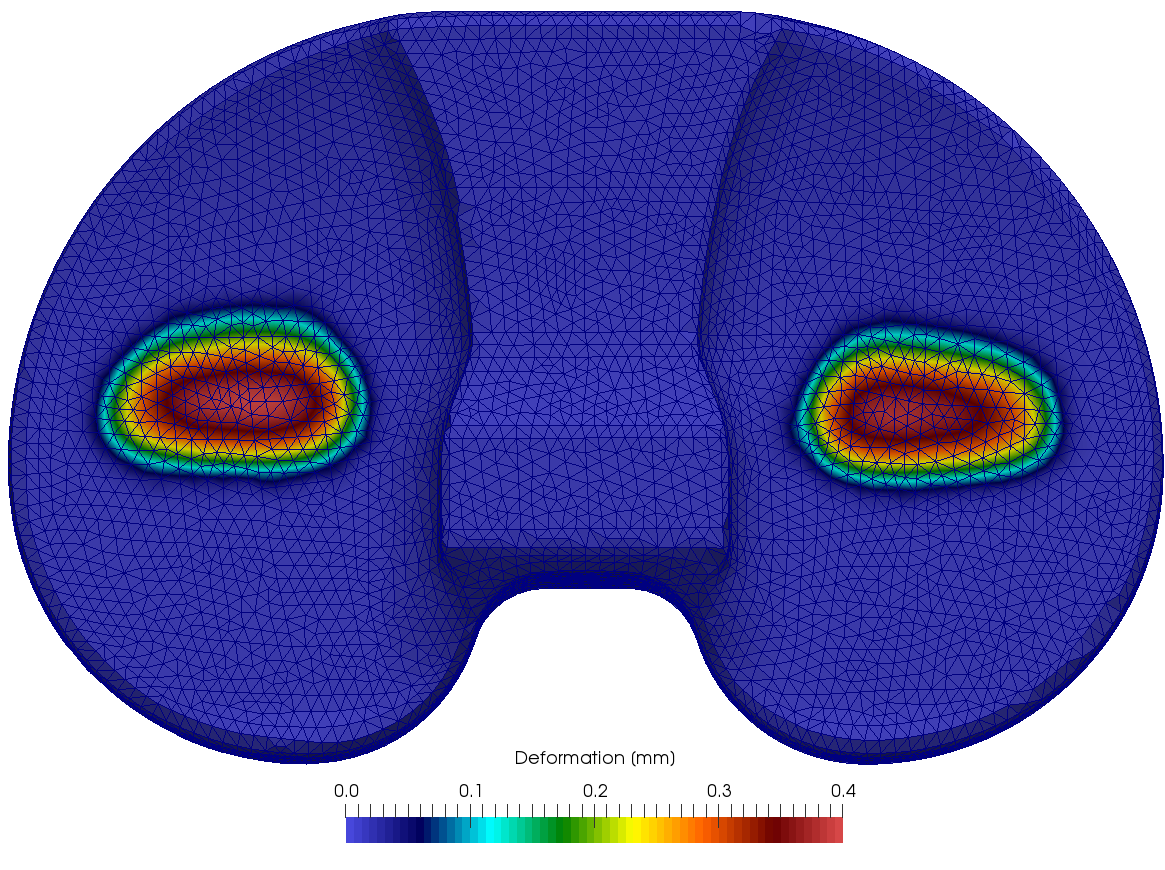}
  \includegraphics[width=0.3\textwidth]{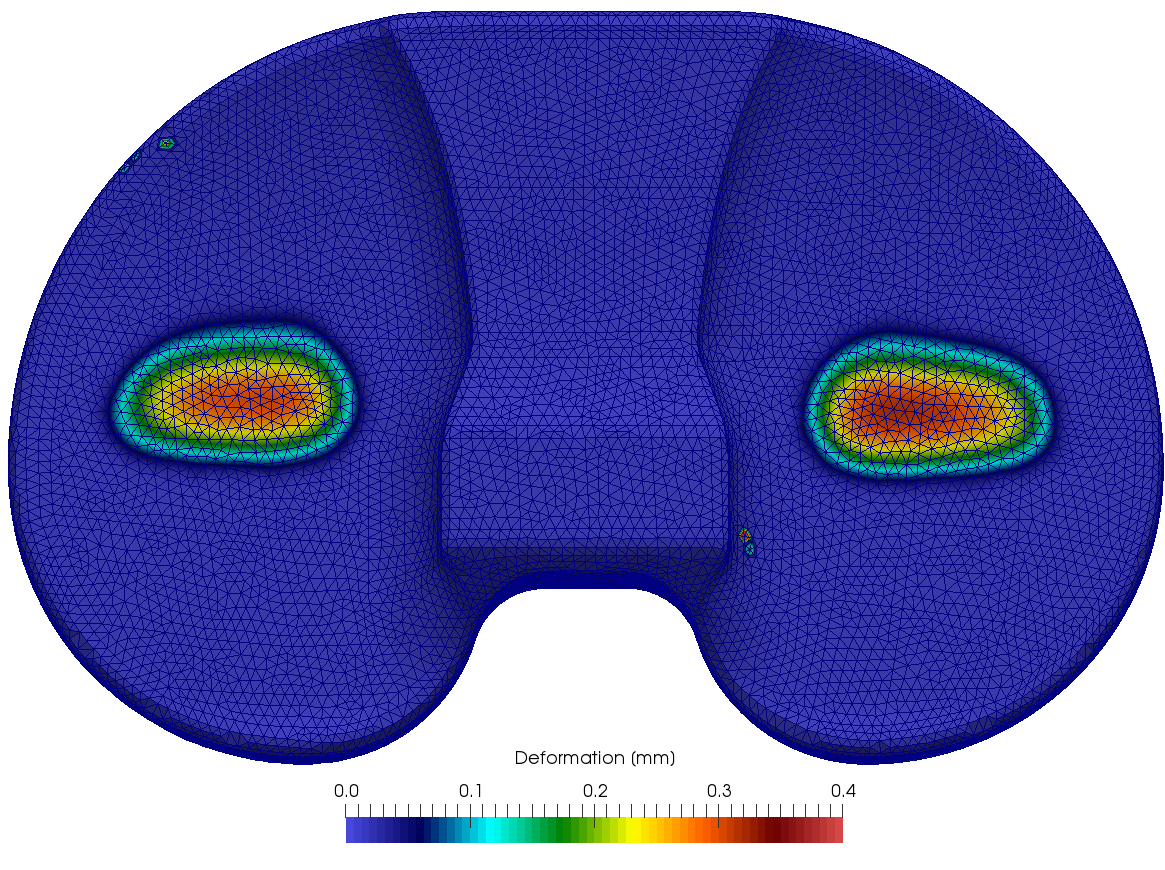}
  \caption{%
    Spatial distribution of wear for the Mebio implant.  Color denotes
    the wear depth.
  }
  \label{fig:wear-area-mebio}
\end{figure}

\begin{figure}
  \centering
  \includegraphics[width=0.3\textwidth]{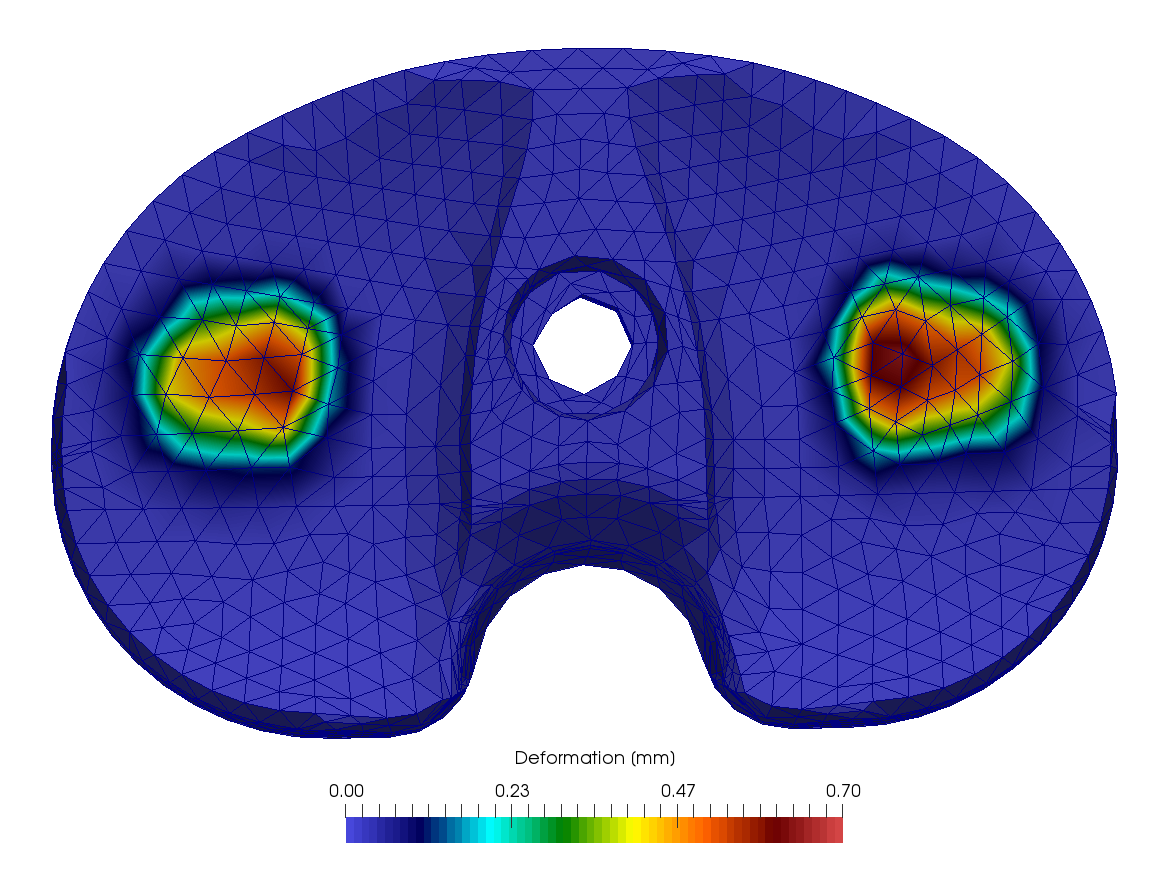}
  \includegraphics[width=0.3\textwidth]{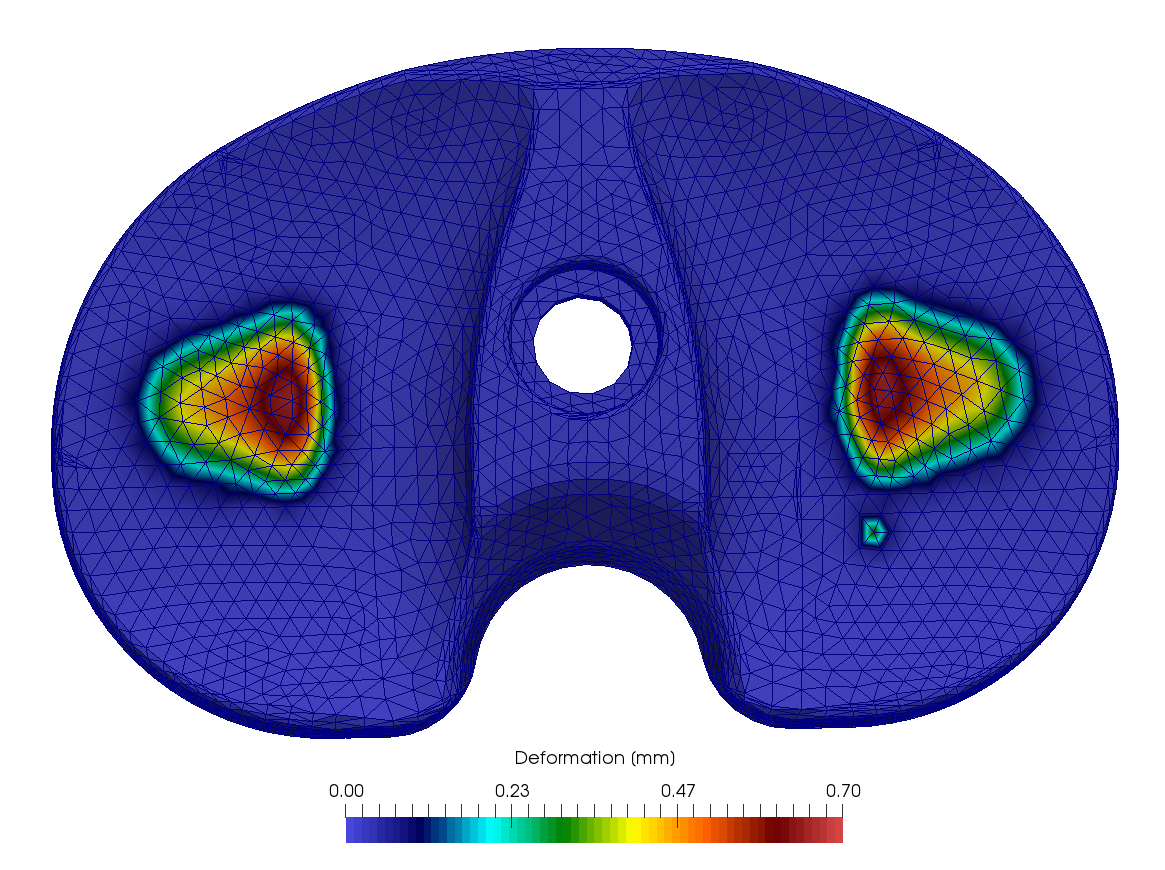}
  \includegraphics[width=0.3\textwidth]{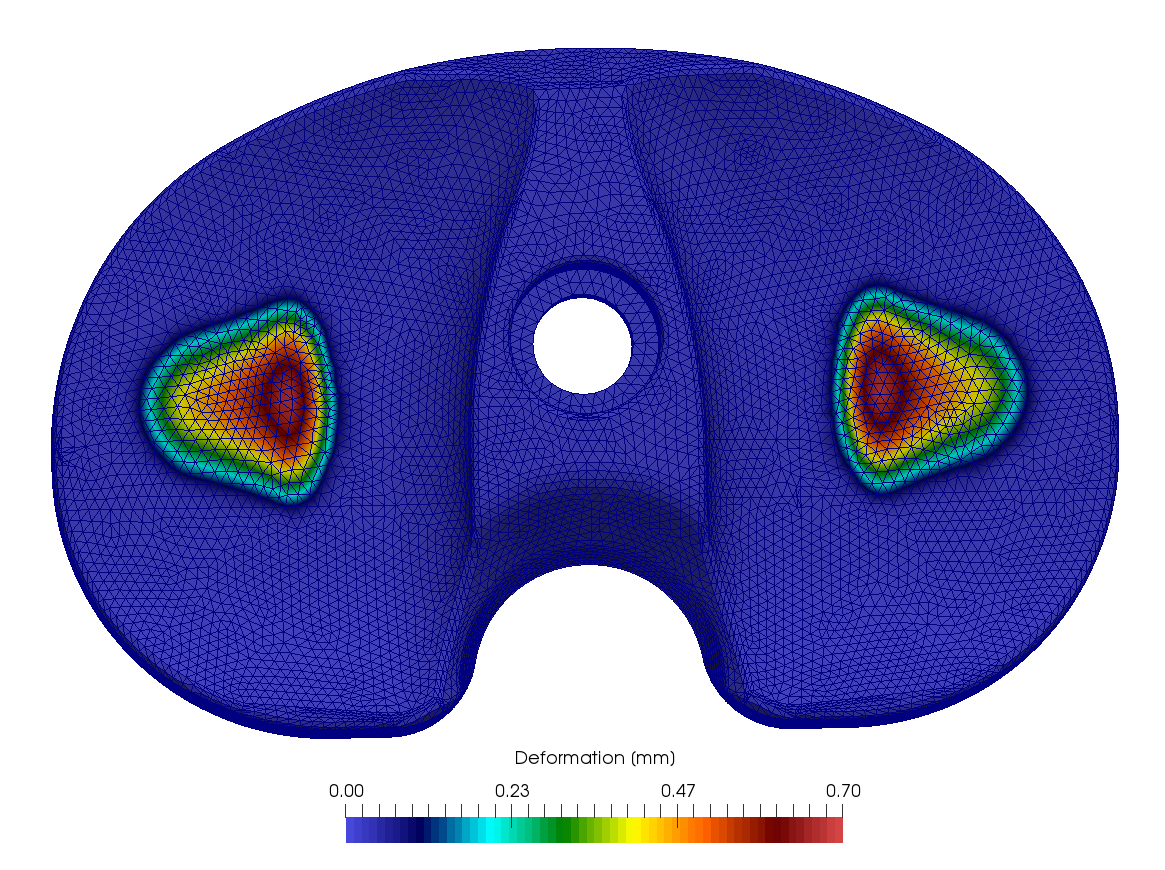}
  \caption{%
    Spatial distribution of wear for the Genius Pro implant. Color denotes
    the wear depth.
  }
  \label{fig:wear-area-geniuspro}
\end{figure}

\begin{figure}
  \centering
  \includegraphics[width=.8\textwidth]{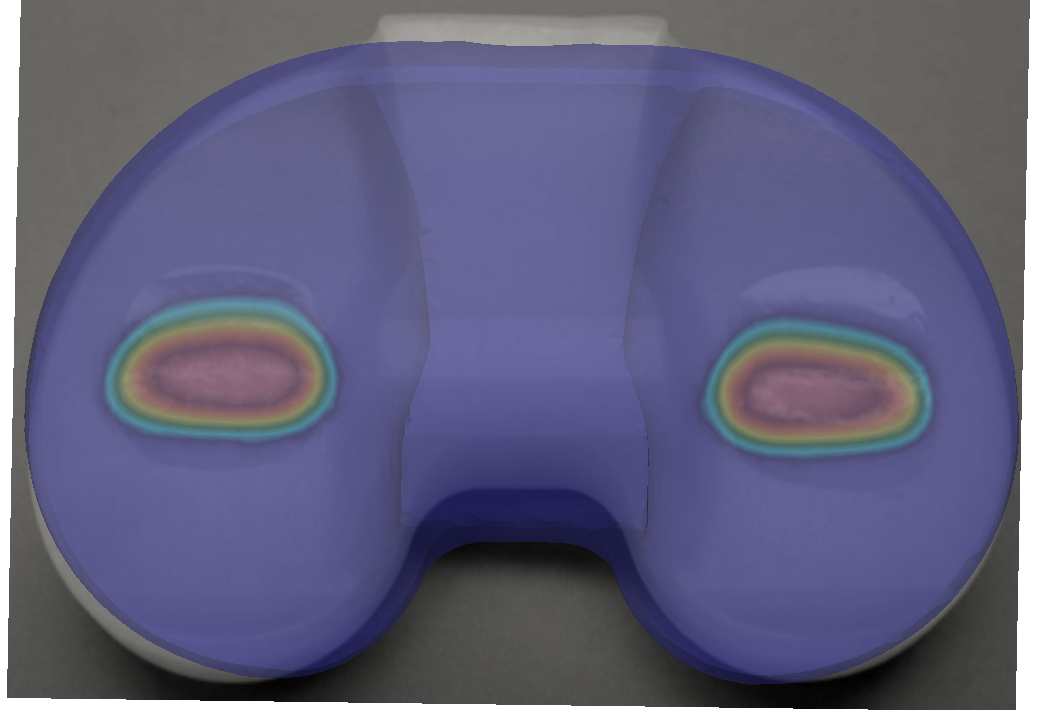}
  \caption{Overlay of simulated and experimental wear for Mebio implant
           after 5~million gait cycles}
  \label{fig:overlayed:mebio}
\end{figure}

Figures~\ref{fig:wear-area-mebio} and~\ref{fig:wear-area-geniuspro} show the simulated wear patterns for both implants.
For the Mebio implant, sizes, shapes and positions of the wear marks match the corresponding
experimental results very well, as can be seen in the overlay in Figure~\ref{fig:overlayed:mebio}.
Wear occurs on 12.5 -- 15.6\% of the proximal side of the tibial inlay in the simulation compared to
13.7 -- 16.5\% in the experiment.  These values were computed by manually segmenting the wear marks,
and counting pixels.

\begin{figure}
  \centering
  \includegraphics[width=.48\textwidth]{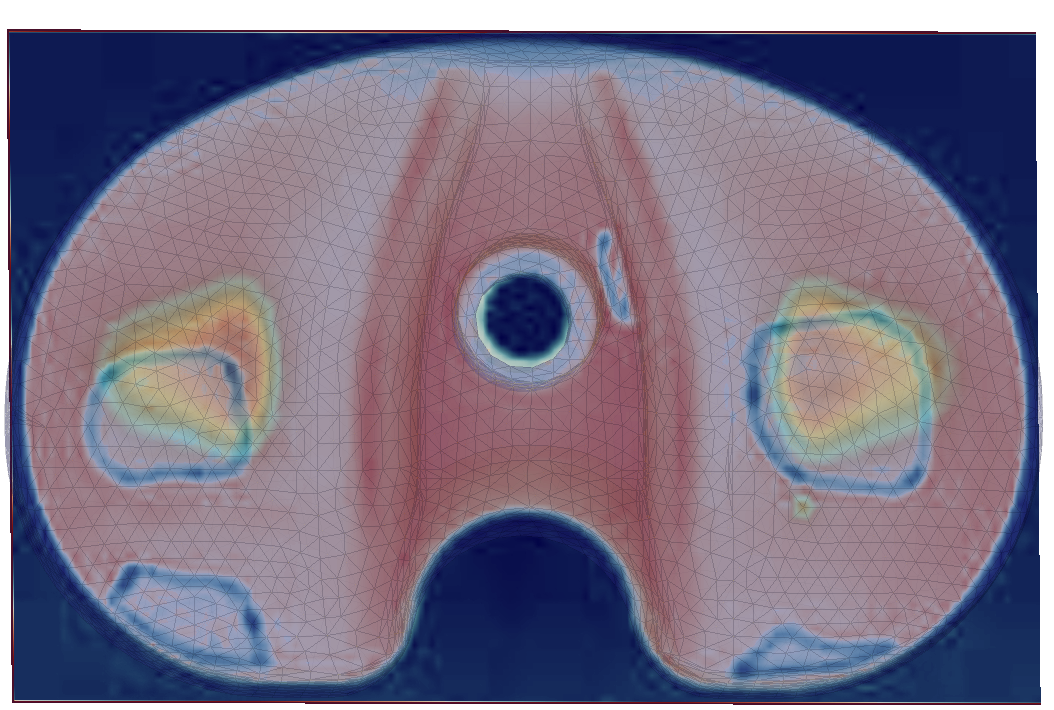}
  \includegraphics[width=.48\textwidth]{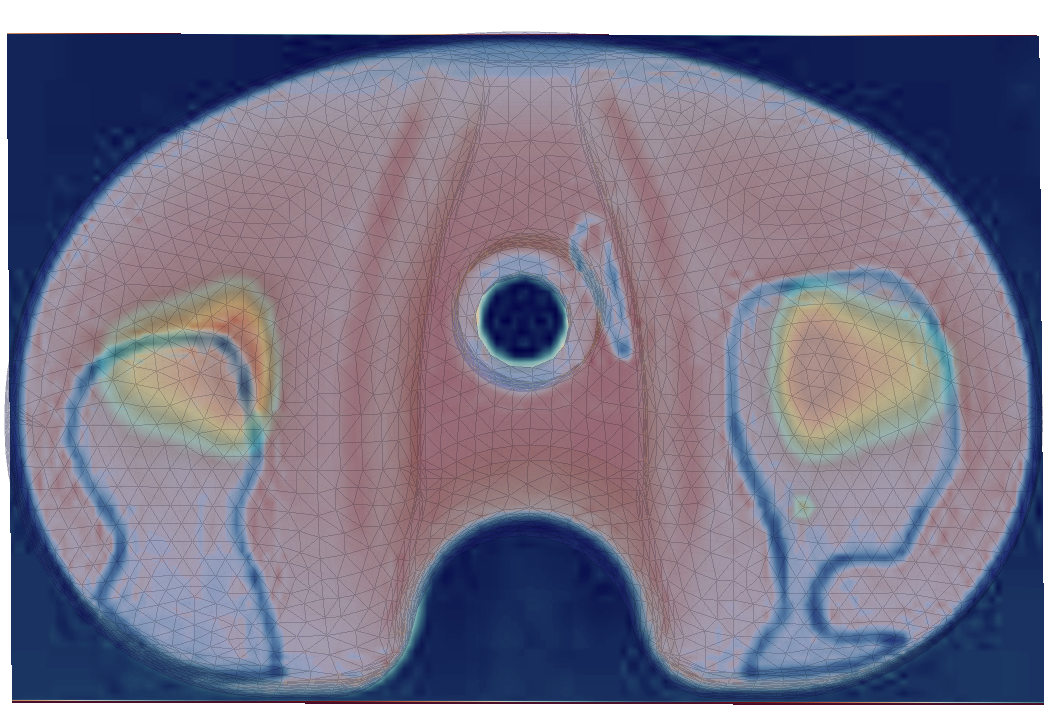}
  \caption{Overlay of simulated and experimental wear for Genius Pro implant
        after 5~million gait cycles. The wear marks on the edges of
        the implement are typical in this type.  They are caused by anterior--posterior
        translations in the swing phase.
  }
  \label{fig:overlayed:geniuspro}
\end{figure}

The experimental wear pattern of the Genius Pro implant shows several secondary
wear marks near the border of the tibial plateau (Figure~\ref{fig:overlayed:geniuspro}).
These are typical for this type of implant, and they are caused by anterior--posterior translations
in the swing phase.  The numerical simulation does not show these marks; presumably a linearization effect.
The main wear marks do not appear quite at the positions of the experimental ones.
Wear occurs on approximately 10\% of the surface in the simulation; on the experimental side the area is much larger (about 20\%).
From the distribution one can see that the numerical simulation does not resolve the increase
of the wear area due to anterior--posterior movement.

In both cases, the experiments show two distinct depressions in the two main wear marks, separated by small
ridges (Figure~\ref{fig:mebio:sample1}).  This effect is not reproduced by the numerical algorithm.

\begin{figure}
  \centering
  \includegraphics[width=.48\textwidth]{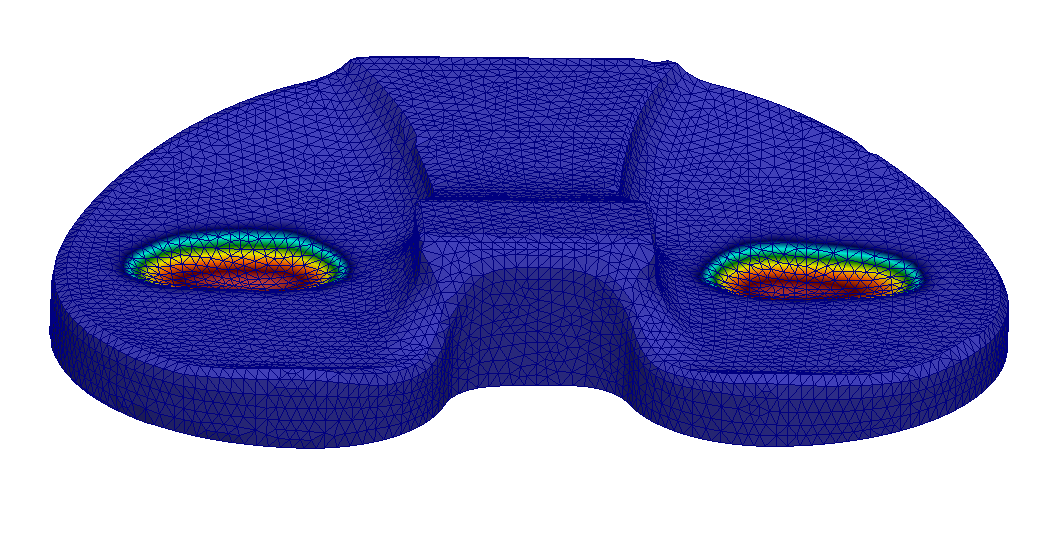}
  \includegraphics[width=.48\textwidth]{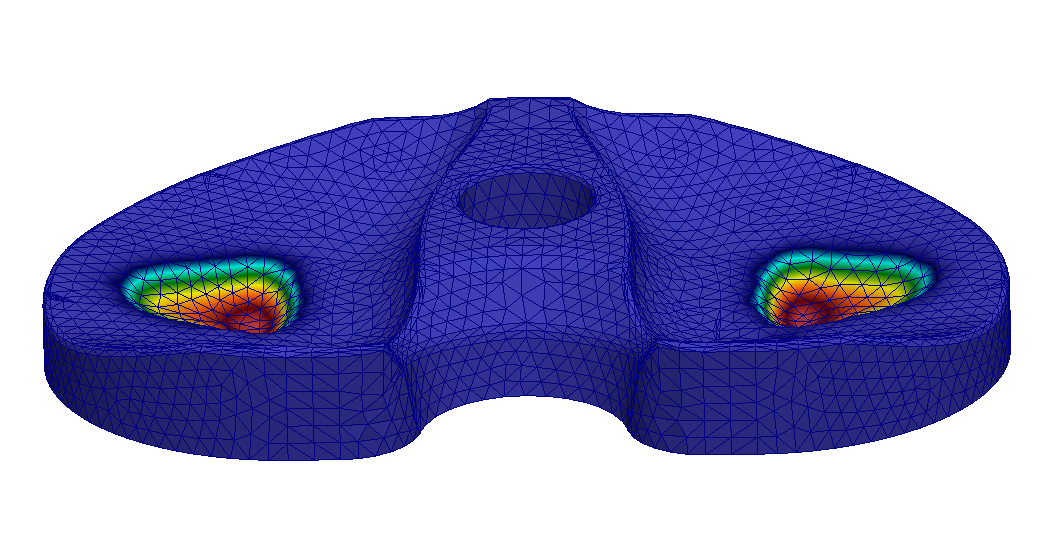}
  \caption{%
    Tibial component of the Mebio (left) and Genius Pro (right) implants deformed by wear after 5~million simulated gait cycles.
    The deformation is scaled by a factor of $5$ to be better visible.
  }
  \label{fig:deformed-tibia}
\end{figure}

The simulation also makes precise predictions about the wear depth at each point on the tibial surface.
Figure~\ref{fig:deformed-tibia} shows the deformed top surface of the tibial inlay after 5~million gait cycles.
The vertical elevation has been stretched by a factor of $5$, to make the wear pattern more visible.

When using a single processor (Intel Xeon E5-2690, 2.9 GHz)
the total simulation on medium-sized meshes with 12 to 14 thousand degrees of freedom in total took 22--30 hours.
By far the majority of time
was spent solving the contact problems.  These problems were solved up to a relative accuracy of $10^{-8}$ for the
displacement fields.  More run-time can be saved when less accuracy is desired.

As the 100 contact problems of a single gait cycle can be computed independently from each other, it is straightforward
to distribute them across several processors.  The contact problems take the majority of the run-time,
and therefore we expect almost linear scaling for up to 100~processors.  Indeed, when repeating the simulations with 10~processors,
the simulation took only between $2.7$ and $3.6$ hours.  This is a speed-up
of about 8.1--8.4 compared to the single-processor run.

\section{Discussion}

The results show that a numerical simulation can produce quantitative results of the wear on a TKR during
an ISO~14243-1 testing cycle that are very close to the results obtained by actual experiments.  Therefore,
while numerical simulations are not yet reliable enough to completely replace in vitro pre-clinical testing,
they nevertheless have the potential to reduce the number of tests in the design and pre-clinical phase,
thereby reducing development costs and time-to-market.

As a particular example, note that the ISO~14243-1 document does not completely specify the initial
mutual positioning and orientation of the femural and tibial TKR parts, because, by necessity,
they are somewhat design-dependent.
However, the initial positioning is something that even the manufacturer needs to determine by experimental testing.
Therefore, occasionally tests need to be aborted, because faulty initial positioning will lead to artifacts
like the wear marks near the circular hole seen in Figure~\ref{fig:overlayed:geniuspro}.
Here, numerical tests can be of great help,
as they allow to cheaply check whether a given position is reasonable.

A standard ISO~14243-1 test simulating 5~million gait cycles takes about three month to complete.
It is no particular challenge to construct finite element models that compute approximate wear marks
and mass loss in less time than that.  Nevertheless, it is desirable to have computer codes that are
as fast as possible.  The less time is needed for an in silico estimation of the wear behavior
of a particular TKR, the better this information can be integrated into the design process.
This is where our finite element model excels.  A complete simulation with $n_\Delta = 5\cdot10^4$
takes only about 22--30~hours on a single
processor.  Even better, it scales almost linearly with the number of processors available (up to 100,
because that is the number of substeps in the ISO~14243-1 gait cycle specification).
As multi-processor machines have become cheap and commonplace over the years, it is no problem to
obtain wear results for a complete test in even under half an hour.
As a further bonus, our
model is guaranteed to produce the result in all situations.  No fine-tune of load-stepping parameters
or similar human intervention is necessary.  The algorithm runs completely autonomous.  This saves
precious human resources.

Furthermore, rapid numerical simulations open new possibilities even early in the design process.
The impact of different variants of the implant shape on the wear behavior can be assessed without
ever constructing a physical prototype.  Modern mathematical techniques known as shape-optimization
even allow to create optimal shapes automatically~\citep{haslinger_maekinen:2003,Willing2009}.
However, such methods can only be used in practice if a fast and reliable simulation tool for
wear is available.

\bigskip

Archard's wear law is the simplest one in a list of different models for mechanical wear.
It is shown to give reasonably accurate results in many situations \citep{Willing2009,OBrien2013}.
Other wear laws try to take into account additional material properties such as cross-shearing,
creep or dependency of the wear factor on contact pressure \citep{Abdelgaied2011,Liu2010,OBrien2014,Turell2003,Strickland2012},
but none of them could be established as a standard so far.
\cite{OBrien2014} compare six different wear models, and obtain good results using Archard's law.

We have found Archard's law to be sufficient to obtain quantitative resuls of very good accuracy
(Section~\ref{sec:results}).  With this goal met,
we can benefit from its advantages.  The most important one is that Archard's law contains only one unknown
parameter, the wear coefficient $k$.  We were able to pick a good parameter value from the
literature, without any recourse to the TKR experimental data available to us.
More precise values of the wear coefficient $k$ also depend on factors like the precise material
and the method used to sterilize it.  However, in production situations this is not a problem,
as $k$ can be measured easily in separate experiments.
More advanced wear laws would require additional parameters, which are in general not easy
to obtain.  Most likely, we would need additional TKR experimental data to determine
material parameters by some form of regression analysis.
Some laws additionally include internal states for which separate differential equations need to be
solved.  With such a wear law we would lose the ability to compute the hundred time steps of
one gait cycle in parallel, sacrificing a lot of computational performance.

\bigskip

\begin{figure}
  \centering
  \includegraphics[width=.8\textwidth]{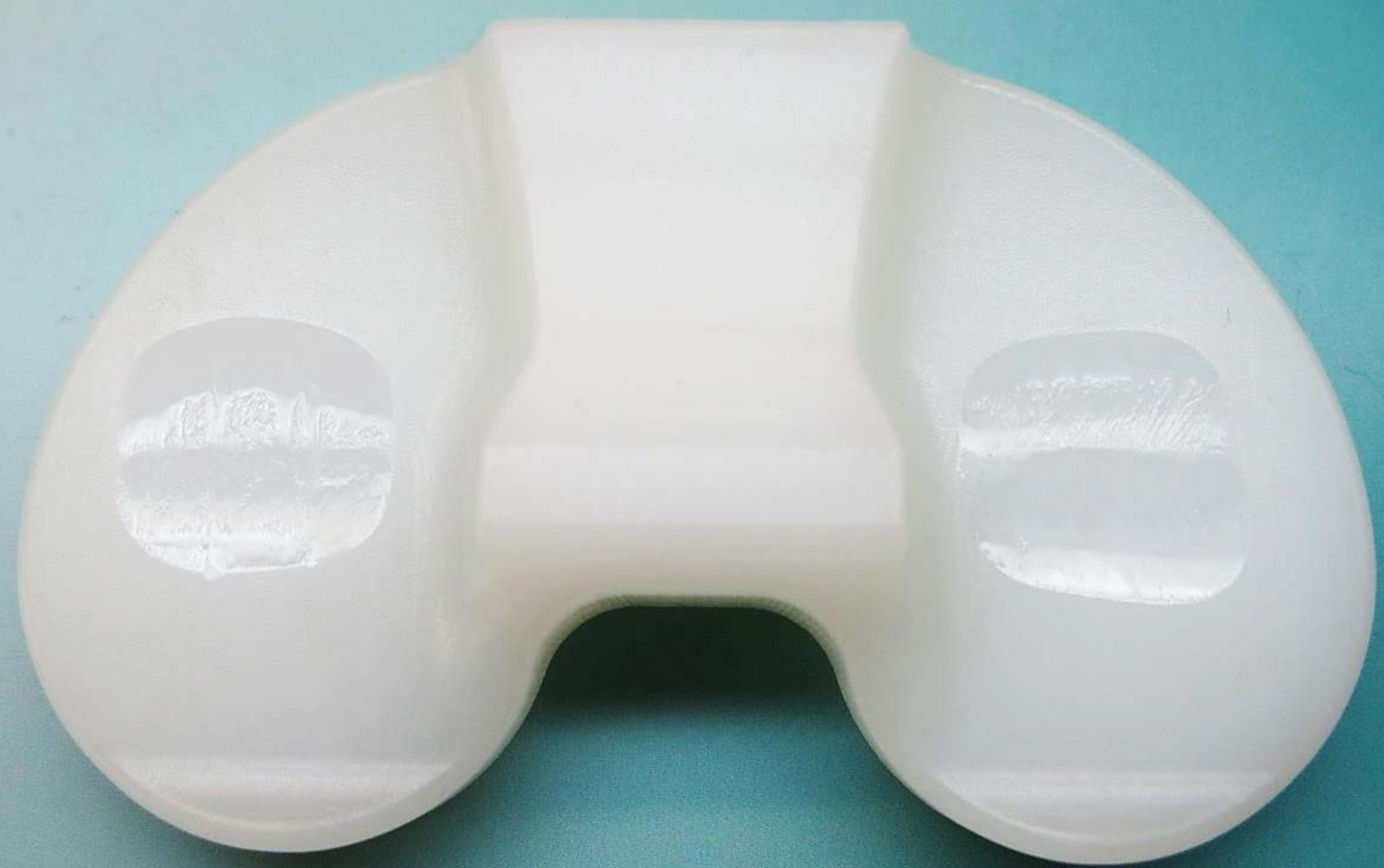}
  \caption{%
    Experimental wear pattern on one of the samples of the Mebio implant.
    The separation in two valleys with a ridge in between can be seen.
  }
  \label{fig:mebio:sample1}
\end{figure}

In in vitro experiments it is frequently observed that the two main wear marks on the tibial plateau
are not simple ``holes''.  Rather, they are typically formed from two or even three individual
depressions, separated by ridges in medial--lateral direction (Figure~\ref{fig:mebio:sample1}).
The origin of this special structure is unclear.  Peculiarities of the testing gait cycle,
friction effects, and plastic behavior of the UHMWPE are all plausible hypotheses.  In our
numerical experiments we have not ever been able to observe the lateral substructure.  This does not
rule out any possibilities, but it suggests that the load pattern by itself is not responsible.
As the nature of this substructure is a practically relevant question, future research will have to
try to capture it with more advanced models.

\section{Conflict of interest statement}

Ansgar Burchardt and Oliver Sander have no conflict of interest to disclose.
Christian Abicht works for Questmed GmbH, an accredited test laboratory for physical testing of implants.

\section{Acknowledgments}

Ansgar Burchardt acknowledges support by the Project ``05M2013--SOAK: Simulation des Abriebs von
Knieimplantaten und Optimierung der Form zur patientengruppenspezifischen Abriebminimierung'',
funded by the German Federal Ministry of Education and Research (BMBF).

\bibliography{burchardt_abicht_sander_wear_trk-bibtex}
\bibliographystyle{elsarticle-harv}

\end{document}